\documentclass[preprint,12pt,authoryear]{elsarticle}
\usepackage{url}
\usepackage{xcolor}
\usepackage{cite}
\usepackage{amsmath,amssymb,amsfonts}
\usepackage{algorithmic}
\usepackage{multirow}
\usepackage{tabularx}
\usepackage{array}
\usepackage{threeparttable}
\usepackage{graphicx}
\usepackage{textcomp}
\usepackage{rotating}

\journal{Medical Image Analysis (under review)}

\begin{document}
\begin{frontmatter}

\title{CATFA-Net:~A Trans-Convolutional Approach for Accurate Medical Image Segmentation}
\author{Siddhartha Mallick$^{a,*,\dagger}$\corref{cor2}, Aayushman Ghosh$^{b,c,*}$\corref{cor2}, Jayanta Paul$^a$, Jaya Sil$^{a,}$\corref{cor1}} 
\affiliation{organization={Department of Computer Science and Technology},
            country={Indian Institute of Engineering Science and Technology, Shibpur 711103, West Bengal, India}}
\affiliation{organization={Department of Electronics and Telecommunication Engineering},
            country = {Indian Institute of Engineering Science and Technology Shibpur 711103, West Bengal, India}}
\affiliation{organization = {Department of Electrical and Computer Engineering}, 
            country = {University of Illinois Urbana-Champaign, Urbana, IL 61801, USA}}
\cortext[cor2]{These authors contributed equally to this work.}
\cortext[cor1]{Corresponding author.}
\cortext[]{Email: {\textcolor{blue}{js@cs.iiests.ac.in}} (J.~Sil); {\textcolor{blue}{aghosh14@illinois.edu}} (A.~Ghosh)}
\cortext[]{$^{\dagger}$Currently with Natwest India}

%% Abstract
\begin{abstract}
Convolutional blocks have played a pivotal role in advancing medical image segmentation by excelling in dense prediction tasks. However, their inability to effectively capture long-range dependencies has limited their performance. Transformer-based architectures, leveraging attention mechanisms, address this limitation by modeling global context and generating expressive feature representations. This potential has been explored in recent researches by introducing hybrid frameworks that integrate transformer encoders with convolutional decoders. Despite their merits these approaches face challenges such as limited inductive bias, high computational overhead, and reduced robustness to data variability. To overcome these limitations, this study proposes CATFA-Net, a novel and efficient segmentation framework designed to deliver high-quality segmentation masks while optimizing parameters, computational cost, and inference speed. CATFA-Net adopts a hierarchical hybrid encoder architecture with a lightweight convolutional decoder backbone. Its transformer-based encoder leverages a novel Context Addition Attention mechanism, which effectively captures inter-image dependencies without incurring the quadratic complexity of standard attention mechanisms. Features from the transformer arm are fused with those from the convolutional arm through a proposed Cross-Channel Attention mechanism, enhancing spatial and channel-dimension information retention during downsampling. Additionally, a Spatial Fusion Attention mechanism in the decoder refines representations while mitigating ambiguous background noise. Comprehensive evaluations on five publicly available datasets—GLaS, DS Bowl 2018, REFUGE, CVC Clinic DB, and ISIC 2018—demonstrate CATFA-Net's superior performance in accuracy and efficiency. The framework sets new state-of-the-art Dice Scores on GLaS {(94.48\%)} and ISIC 2018 {(91.55\%)}. Robustness analyses and external validation further establish its strong generalization capabilities in binary segmentation tasks. \\ 
{Code: \url{https://github.com/aayushmanghosh/CATFA-Net-pytorch.git}}
\end{abstract}

%%Graphical abstract
%\begin{graphicalabstract}
%\includegraphics{grabs}
%\end{graphicalabstract}

%%Research highlights
%\begin{highlights}
%\item Research highlight 1
%\item Research highlight 2
%\end{highlights}

%% Keywords
\begin{keyword}
Medical image segmentation \sep Transformers \sep Convolution blocks \sep Self Attention \sep Hybrid Attention
\end{keyword}
\end{frontmatter}

%% main text
% Introduction
\section{Introduction}
Creating reliable techniques for organ isolation and abnormality detection based on scanned images remain a critical challenge in medical imaging \textcolor{blue}{\citep{ma2024segment, shaker2024unetr++,mallick2023response,chen2021transunet,Wang_2022,wang2019deeply}}. These tasks are fundamental for computer-aided diagnosis and play a vital role in helping clinicians in planning precise surgical interventions. Over the years, various machine learning approaches have been developed for organ segmentation with deep convolutional networks emerging as leading candidates due to their powerful nonlinear learning capabilities. Architectures such as U-Net \textcolor{blue}{\citep{ronneberger2015unet}}, UT-Net \textcolor{blue}{\citep{gao2021utnet}}, V-Net \textcolor{blue}{\citep{milletari2016vnet}}, Dri-Net \textcolor{blue}{\citep{chen2018drinet}}, U-Net++ \textcolor{blue}{\citep{zhou2018unet++}}, R2-UNet \textcolor{blue}{\citep{alom2018recurrent}}, and Dense-UNet \textcolor{blue}{\citep{denseunet}} have demonstrated exceptional performance in both image and volumetric segmentation across diverse medical imaging modalities, highlighting the efficacy of {Convolutional Networks (ConvNets)} in extracting discriminative features for organ or lesion segmentation in medical scans. Despite these advancements, ConvNets struggle to capture global and long-range dependencies, as they primarily focus on local patterns within pixel subsets. Approaches such as residual networks \textcolor{blue}{\citep{alom2018recurrent}}, atrous convolutions \textcolor{blue}{\citep{weightedresunet}}, and attention mechanisms \textcolor{blue}{\citep{shi2022agnet, chen2022aau, valanarasu2021medical}} have attempted to address this limitation, but nonetheless, it is important to acknowledge that there remains substantial potential for improvement. 

\begin{figure}[t]
    \centering
    \includegraphics[width=1.0\linewidth]{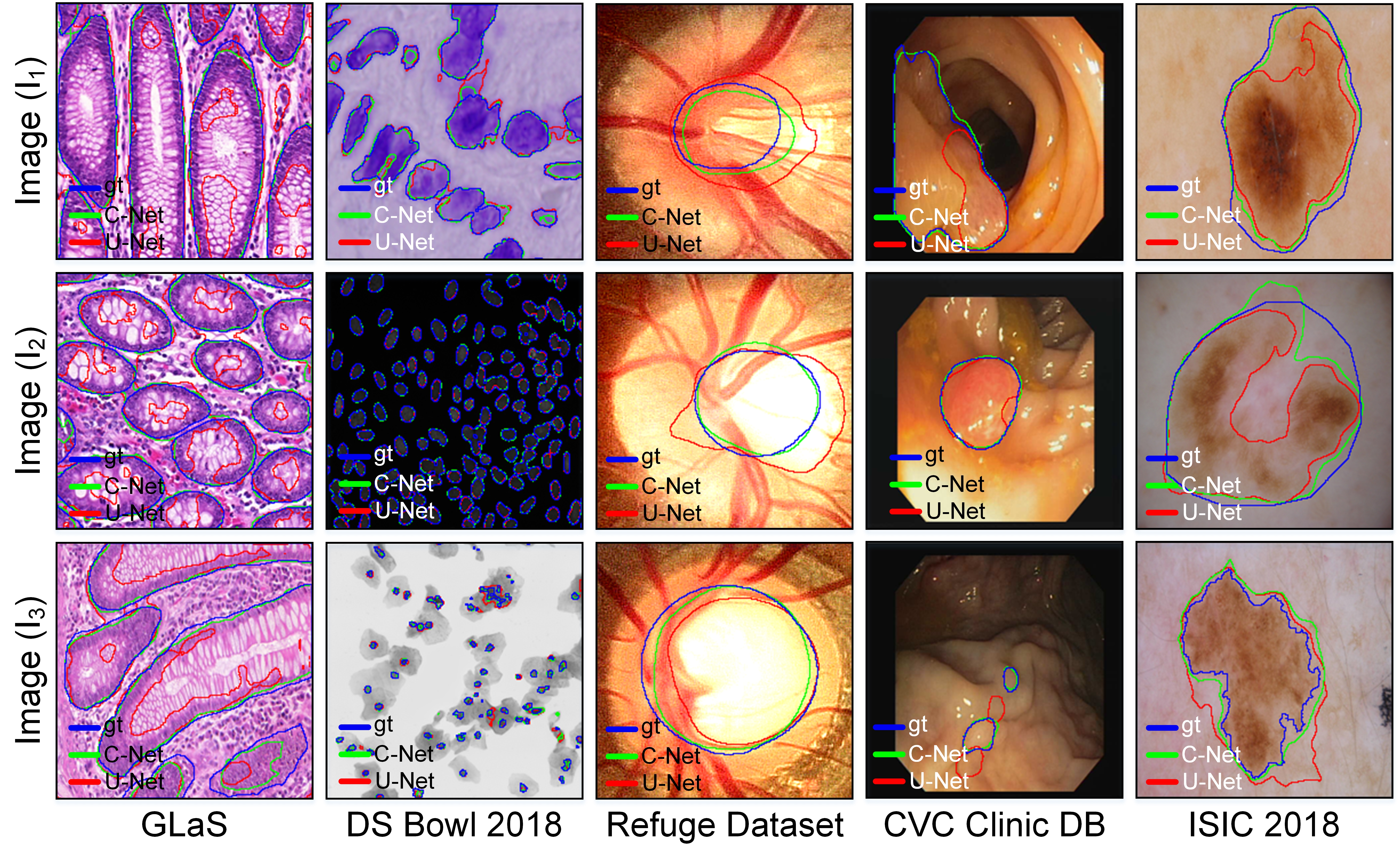}
    \caption{\textbf{Demonstrating the importance of modeling long-range dependencies.} Examples from various medical imaging benchmarks (GLaS, DS Bowl 2018, REFUGE, CVC Clinic DB, ISIC 2018) are shown. Blue outlines represent the ground truth (gt), red outlines indicate U-Net predictions, and green outlines show predictions from CATFA-Net (C-Net). While the convolution-based U-Net method misclassifies regions in several datasets due to its limited ability to capture long-range dependencies, CATFA-Net effectively addresses this limitation.}
    \label{figure-1}
\end{figure}

To first understand the importance of incorporating long-range dependencies in segmentation tasks, {we analyze} medical images from several well-established public datasets (see Fig. \ref{figure-1}, and {section 3.1}), highlighting the limitations of convolutional networks such as U-Net. As {shown in Fig. 1, across images from different datasets, U-Net often misclassifies background regions as part of the target organ. This can be attributed to its limited capacity to model long-range dependencies between spatially distant pixels.} Conversely, methods capable of learning long-range dependencies-such as the proposed {CATFA-Net} framework—excel by accurately distinguishing between mask pixels and background regions. Additionally, in cases where the segmentation mask is extensive (see column labeled GLaS in Fig. \ref{figure-1}) or background patterns are complex, understanding long-range dependencies among pixels within the mask can further enhance the accuracy of predictions. 

In natural language processing (NLP), transformers have proven highly effective at capturing global context by identifying dependencies within sequential input using multi-head self-attention (MSA) \textcolor{blue}{\citep{vaswani2017attention}}. Inspired by their success, transformers have been adapted for computer vision tasks \textcolor{blue}{\citep{dosovitskiy2021image, liu2021swin, xie2021segformer, carion2020end}}. However, their quadratic complexity with respect to input size presents challenges for medical imaging applications. While several studies have sought to optimize self-attention mechanisms for efficiency \textcolor{blue}{\citep{child2019generating, wang2020linformer, maaz2022edgenext, kitaev2020reformer}}, focus was mostly given on classification tasks, leaving dense prediction tasks under-explored. Additionally, transformers require large-scale datasets for optimal performance \textcolor{blue}{\citep{dosovitskiy2021image}} and exhibit low inherent bias, making them less effective when trained on small, label-limited medical datasets. By contrast, convolutional operations offer higher inductive bias and perform well with limited data. Nevertheless, the contributions of transformer in modeling global relations from medical image cannot be rendered insignificant. Therefore, a hybrid architecture that leverages both approaches is essential to advancing medical imaging segmentation performance. Recent works, including transformer-based encoders with convolutional decoders \textcolor{blue}{\citep{chen2021transunet, zhang2021transfuse, ji2021multi}} and hybrid designs for both encoder and decoder networks \textcolor{blue}{\citep{hatamizadeh2021unetr, valanarasu2021medical, zhou2023nnformer}}, have improved segmentation accuracy but often result in high computational demands and reduced robustness. These limitations are attributed to inefficient self-attention designs and unnecessary architectural complexity. 

The proposed CATFA-Net addresses these challenges by introducing an efficient hybrid hierarchical trans-convolutional framework designed to enhance both segmentation accuracy and operational efficiency in terms of {trainable parameters, memory footprint, robustness, and inference speed.} CATFA-Net is structured with two distinct encoder stages: a convolutional arm and a transformer arm, complemented by a series of convolutional blocks in the decoder stage. Central to the design is the hierarchical context addition transformer (H-CAT) arm, which integrates (1) a context addition self attention mechanism for  efficient inter-image dependency modeling without the quadratic complexity, and (2) depth-wise fully convolutional layers to enhance performance while effectively removing positional information. The outputs from each of the encoder arms are fused using a proposed Cross-Channel Trans-Convolutional Fusion Attention mechanism, preserving critical spatial and channel-dimension information. Then the integrated output from each stage of the encoder arm is subsequently connected to the decoder stage via skip connection and a proposed spatial attention fusion gate that reduces the ambiguous background noise and emphasizes salient features during up-sampling. Lightweight convolutional blocks complete the up-sampling process, culminating in an architecture that improves segmentation accuracy and efficiency. Evaluations demonstrate that CATFA-Net significantly enhances performance on multiple medical imaging benchmarks while addressing critical issues of computational overhead and robustness. {The remainder of the paper is organized as follows. Section II details the design of the proposed segmentation network, including the rationale behind key architectural choices. Section III outlines the implementation settings, followed by quantitative and qualitative comparisons with baseline models on multiple public medical imaging datasets, along with ablation studies and a robustness analysis of the proposed framework. Section IV presents a discussion of the experimental results, highlights key findings, and outlines the limitations of the current study. Finally, Section V concludes the paper.}

\begin{figure}
    \centering
    \includegraphics[width=1\linewidth]{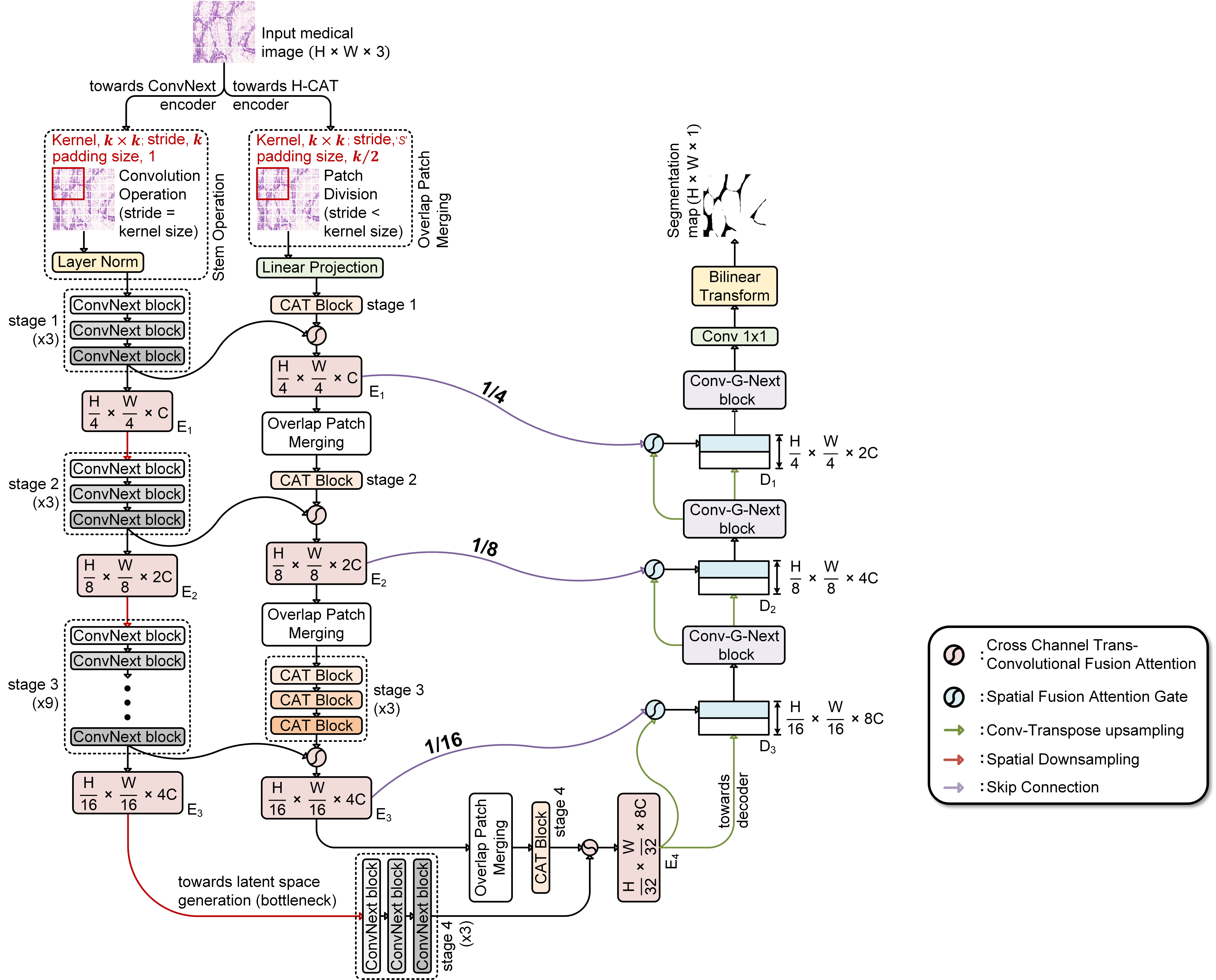}
    \caption{\textbf{Overview of the proposed CATFA-Net model for efficient medical image segmentation}. This architecture integrates ConvNeXt and H-CAT encoder branches for efficient feature extraction, utilizing advanced attention mechanisms such as context-addition attention, which captures inter-image resemblance to enhance feature representation, and cross-channel attention, which maintains consistency across spatial and channel dimensions. The encoder processes the input image through several stages, progressively reducing spatial resolution while deepening feature representation. The decoder reconstructs the segmentation map using spatial fusion mechanisms to reduce background ambiguity, along with bilinear up-sampling and skip connections to preserve spatial context effectively.}
    \label{figure-2}
\end{figure}

%Architecture
\section{Architecture}
As illustrated in Fig. \ref{figure-2}, CATFA-Net is structured around two main components: (1) a trans-convolutional encoder featuring an advanced multi-level feature fusion mechanism, and (2) a lightweight {Conv-G-NeXt} decoder that efficiently integrates and upscales multi-level features to generate semantic segmentation maps. The trans-convolutional encoder itself is split into two separate branches: (1) the {ConvNeXt} encoder branch and (2) the Hierarchical Context Addition Transformer (H-CAT) encoder branch. Each branch processes the input feature map $\mathbf{x}\in \mathbb{R}^{H\times W\times3}$ independently at each encoder stage $s \in {1, 2, 3, 4}$. Notably, the output from the ConvNeXt branch at each stage is passed to the H-CAT branch via the proposed cross-channel trans-convolutional fusion attention (CCTFA) modules.

The H-CAT {encoder is inspired by the hierarchical stage-wise design of the Swin Transformer architecture \textcolor{blue}{\citep{liu2021swin}} and begins with an overlap patch merging layer that partitions} $\mathbf{x}$ into $p \times p$ patches and projects them into $C$ channels, {yielding feature maps $\mathbf{E}_s \in \mathbb{R}^{\frac{H}{p} \times \frac{W}{p} \times 2^{s-1} \cdot C}$, where $p = 2^{s+1}$ for stages 1 through 4. This is effectively a `$2\times$' downsampling step implemented via a strided convolution with kernel size $k$, stride $S$ ($S < k$), and padding $k/2$. In our implementation, $k=3$ and $S=2$ are used in stages 2–4 to preserve local continuity in $\mathbf{E}_s$. Stage 1 uses a larger kernel ($k=7$, $S=4$) and includes linear embedding to map the input into $\frac{H}{4} \times \frac{W}{4} \times C$. The extracted features are then refined by successive CAT blocks and CCTFA modules. In parallel, the ConvNeXt encoder branch applies non-overlapping convolution (kernel $4$, stride $4$) for patch embedding at stage 1 and continues with standard downsampling and ConvNeXt blocks in the remaining stages.}

{To preserve spatial information and suppress background noise, CATFA-Net employs skip connections along with a proposed spatial attention fusion gate between the encoder and decoder pathways. The decoder mirrors the encoder’s four-stage layout, where each stage upsamples the input feature map via transposed convolution, followed by a Conv-G-NeXt block (except the final stage). The channel width is halved progressively across decoder stages. Finally, the output from the last decoder stage passes through a point-wise convolution block followed by bilinear upsampling \textcolor{blue}{\citep{lin2017bilinear}}, which generates the final pixel-wise prediction mask $y \in \mathbb{R}^{H \times W \times 1}$. The following subsections elaborate on the internal design of the CAT blocks, CCTFA modules, and Conv-G-NeXt components.} 

\begin{figure}
    \centering
    \includegraphics[width=1.0\linewidth]{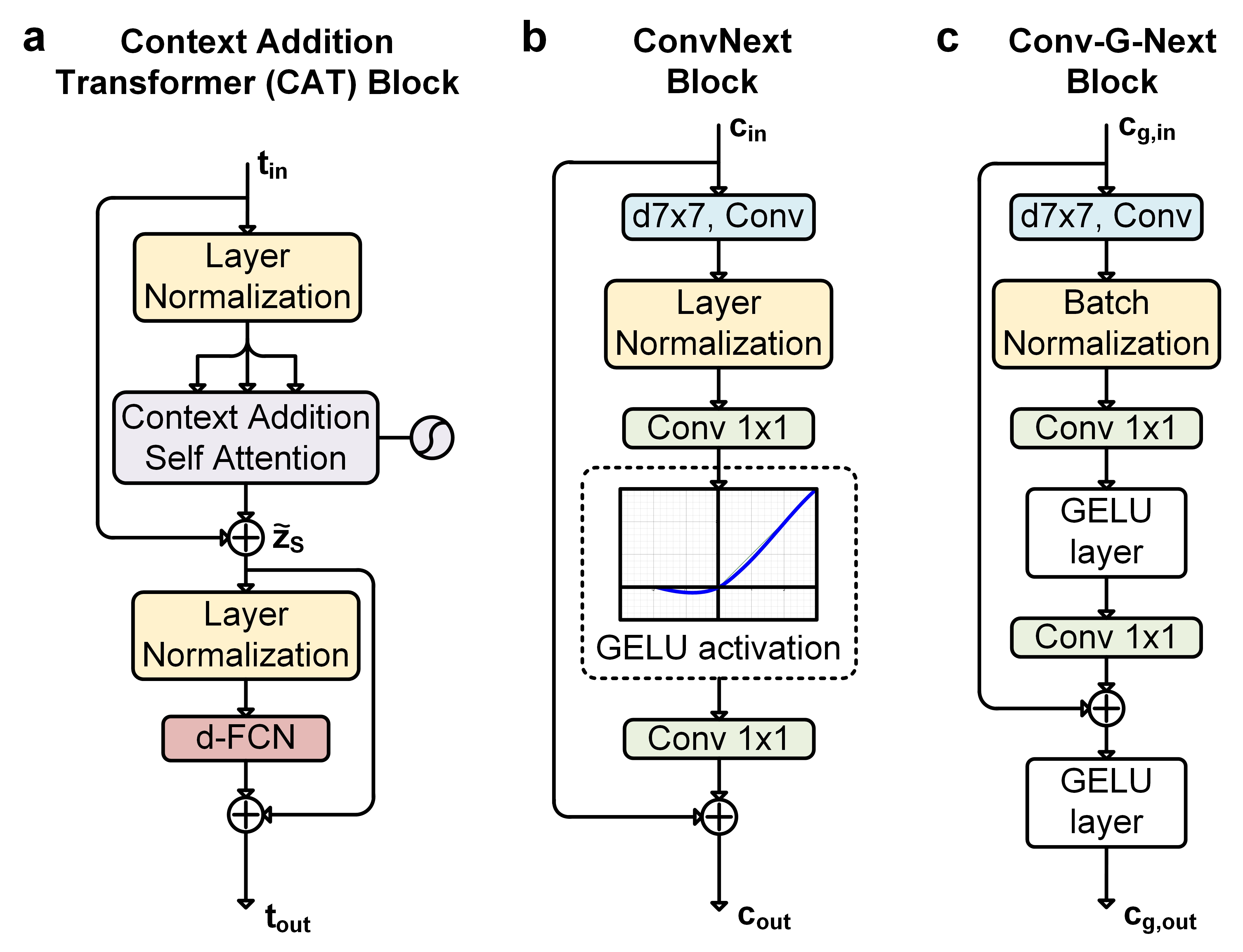}
    \caption{\textbf{Overview the of key building blocks used in the proposed method. (a)} The Context Addition Transformer (CAT) block integrates a novel Context Addition Self-Attention module, LN, and a depthwise Fully Convolutional Network (d-FCN) to capture global and contextual dependencies effectively while learning inter-image relations. \textbf{(b)} The ConvNext block, employing a $7\times7$ convolution kernel, LN, and a pointwise $1\times1$ convolution, provides a lightweight yet powerful local feature extraction mechanism. \textbf{(c)} The Conv-G-Next block built on ConvNext by incorporating Batch Normalization, a GELU activation layer, and an additional $1\times1$ convolution layer, enhancing non-linear transformations and enabling finer-grained feature representation while up-sampling.}
    \label{figure-3}
\end{figure}

\subsection{Context Addition Transformer (CAT) Block}
{Each intermediate feature map $\mathbf{E_s}$ produced by the H-CAT encoder is sequentially processed through multiple CAT blocks (see Figs. \ref{figure-2}, \ref{figure-3}(a)), with the number of blocks determined by the encoder stage $s$. In each CAT block, the standard multi-head self-attention mechanism of the Swin Transformer (Swin-T) is replaced by a context addition self-attention module, specifically designed to model inter-image resemblance. This attention module is followed by a Layer Normalization (LN) layer and a depthwise Fully Convolutional Network (d-FCN) block, which substitutes the Multilayer Perceptron (MLP) found in traditional Swin-T blocks.}
 
\subsubsection{Context Addition Self-Attention}
Consider an intermediate feature map $\mathbf{E_s}$ with dimensions $H/p$, $W/p$, and channels $2^{s-1}\cdot C$ where $p=2^{s+1}$. The self-attention layer computes its output using projected input as follows: 
\begin{equation}
    \text{Attention}(\mathbf{Q},\mathbf{K},\mathbf{V}) = \mathcal{S}\left(\frac{\mathbf{Q} \mathbf{K}^T}{\sqrt{d_{\mathbf{QK}}}}\right)\mathbf{V}
\end{equation}
Here, $d_{\mathbf{QK}}$ is the dimension of the \textit{query}, $\mathbf{Q = W_{Q}E_s}$ and \textit{key}, $\mathbf{K = W_{K}E_s}$ heads, and $\mathcal{S}(\cdot)$ represents the Softmax function. The projection matrices $\mathbf{W_Q,W_K,W_V}$ are learnable parameters. The self-attention mechanism pools values $\mathbf{V}$ based on global affinities computed via $\mathcal{S}(\mathbf{Q} \mathbf{K}^T)$, allowing the capture of non-local information across the entire feature map. However, this process has a computational complexity of $\mathcal{O}(N^2)$,  making it infeasible for high-resolution inputs. Moreover, traditional multi-head self-attention (MSA) often overlooks the global context inherent in adjacent key-query pairs, which is particularly detrimental for medical image datasets characterized by strong inter-image resemblance. 

We propose a two-step approach to address these challenges. First, a Context Attention Pre-attention (CAP) module is introduced as a pre-processing step to enhance the learning of inter-image dependencies. Second, a spatial reduction block is incorporated, reducing computational complexity from $\mathcal{O}(N^2)$ to $\mathcal{O}(N^2/\mathcal{R})$ where $\mathcal{R}$ is the reduction ratio. As illustrated in Figure 4, the CAP module enriches the key representations $\mathbf{K}$ by concatenating $\mathbf{K}$ and $\mathbf{Q}$, followed by a series of $1\times1$ convolutions that capture cross-image resemblances. The updated key $\mathbf{K'}$ is given by:
\begin{equation}
    \mathbf{K'} = \Gamma((\mathbf{K}\oplus\mathbf{Q})\mathbf{W_1^{1 \times 1}}) \cdot \mathbf{W_2^{1\times1}} + \mathbf{K}
\end{equation}
Here, $\mathbf{W_1^{1\times 1}}$ and $\mathbf{W_2^{1\times 1}}$ are weight matrices for the convolution layers; $\Gamma(\cdot)$ is the GELU activation layer and $\oplus$ denotes concatenation. The modified key $\mathbf{K'}$ and the value $\mathbf{V}$ are  then processed through a spatial reduction block, which involves a reshape-transfer function and a linear layer. The resulting dimensions for $\mathbf{\hat{K}}$, and $\mathbf{\hat{V}}$ are $H/(l\cdot R) \times W/(l\cdot R) \times 2^{s-1} \cdot C$.

As illustrated in Fig. 4, the CAP module enriches the key representations $\mathbf{K}$ by concatenating $\mathbf{K}$ and $\mathbf{Q}$, followed by a series of $1\times1$ convolutions that capture cross-image resemblances. The updated key $\mathbf{K'}$ is given by:
\begin{equation}
    \mathbf{K'} = \Gamma((\mathbf{K}\oplus\mathbf{Q})\mathbf{W_1^{1 \times 1}}) \cdot \mathbf{W_2^{1\times1}} + \mathbf{K}
\end{equation}
Here, $\mathbf{W_1^{1\times 1}}$ and $\mathbf{W_2^{1\times 1}}$ are weight matrices for the convolution layers; $\Gamma(\cdot)$ is the GELU activation layer and $\oplus$ denotes concatenation. The modified key $\mathbf{K'}$ and the value $\mathbf{V}$ are  then processed through a spatial reduction block, which involves a reshape-transfer function and a linear layer. The resulting dimensions for $\mathbf{\hat{K}}$, and $\mathbf{\hat{V}}$ are $H/(l\cdot R) \times W/(l\cdot R) \times 2^{s-1} \cdot C$. Across all experiments we set $\mathcal{R}$ to be $\{8,4,2,1\}$ for the four encoder stages, respectively. 

The refined attention heads are then projected through an MLP layer. The outputs from the MLP are concatenated and passed through a linear layer to generate the final output $\mathbf{z'_s}$:
\begin{equation} 
    \mathbf{z'_s} = \mathcal{L}\left(\text{Concat}(\mathbf{h_0}, \mathbf{h_1}, \cdots, \mathbf{h_M})\mathbf{W^O}, 2^{s-1}\cdot C\right)
\end{equation}
where, $\mathbf{h_j} = \text{Attention}(\mathbf{Q,\hat{K},\hat{V}})$ is computed using equation (1), where $j \in \{0,1,\cdots,\mathbf{M}\}$, $\mathcal{L}(\cdot)$ denotes the linear layer, $\mathbf{W^O} \in \mathbb{R}^{C'\times C'}$ represents the parameters of the MLP and linear layers with $C' = 2^{s-1}\cdot C$.   

\begin{figure}
    \centering
    \includegraphics[width=1\linewidth]{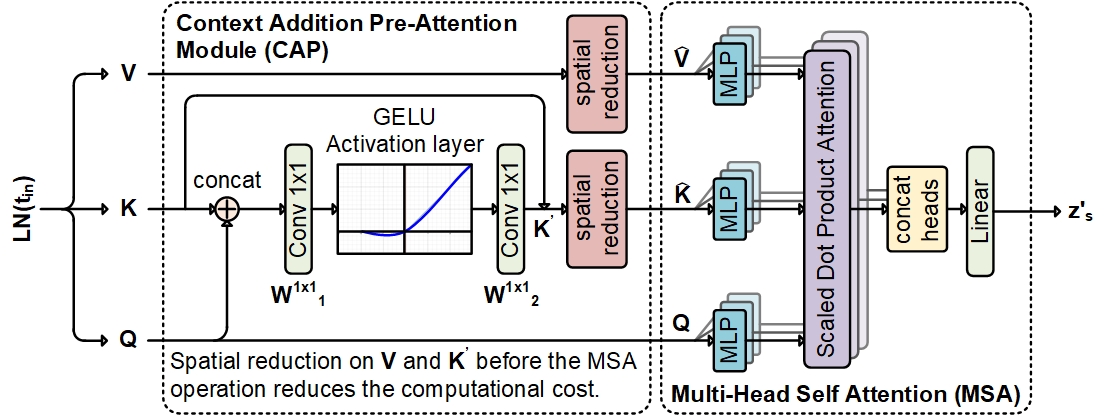}
    \caption{\textbf{Overview of the proposed Context Addition Self Attention block.} The Context Attention Pre-attention (CAP) module enhances the $\mathbf{K}$ bit by using $\mathbf{Q}$ and $1\times1$ convolutions along with GeLU nonlinearity to learn inter-image dependencies. Simultaneously, $\mathbf{V}$ along with $\mathbf{K'}$ is passed through a spatial reduction block, reducing computational complexity of the whole process from $\mathcal{O}(N^2)$ to $\mathcal{O}(N^2/\mathcal{R})$, where $\mathcal{R}$ is the reduction ratio. The modified heads are then processed through a standard MSA block to produce the output $\mathbf{z'_s}$.}
    \label{figure-4}
\end{figure}

\subsubsection{depthwise-Fully Convoluted Network (d-FCN)}
In contrast to convolutional layers, self-attention mechanisms {do not inherently encode} positional information, which is critical for capturing the structural characteristics of objects in vision tasks.~{To address this limitation, various strategies have been proposed to incorporate positional bias into attention-based models.} For instance, the Vision Transformer \textcolor{blue}{\citep{dosovitskiy2020image}} introduces positional encodings (PEs) into input patches to provide locational context. However, the fixed resolution of these encodings creates challenges, as differences between training and testing data require interpolation, potentially leading to reduced accuracy. {To mitigate this issue, several alternatives have been developed, such as data-driven PEs \textcolor{blue}{\citep{chu2021conditional}} and methods that directly embed relative positional bias into the query, key, and value projections \textcolor{blue}{\citep{valanarasu2021medical}}. Interestingly, SegFormer \textcolor{blue}{\citep{xie2021segformer}} challenges the necessity of positional encodings altogether in semantic segmentation, instead introducing Mix-FFN, which leverages zero-padding to encode spatial structure implicitly. Motivated by this insight,} we propose a depthwise fully convolutional network (d-FCN) module that improves performance while effectively encoding positional information through local spatial context.

{As illustrated in Fig. \ref{figure-3}(a), the d-FCN operates on the output $\mathbf{z'_s}$ from the preceding context addition self-attention block, and the intermediate representation $\mathbf{t_{in}}$ from the overlap patch merging stage, which is further normalized via a LayerNorm (LN) layer.~The normalized tensor is then processed through a sequence comprising a $1\times1$ convolution ($\mathbf{W_1^{1\times1}}$), followed by a $3\times3$ depthwise convolution ($\mathbf{W_1^{3\times3}}$), a GELU activation $\Gamma(\cdot)$, and a final $1\times1$ convolution ($\mathbf{W_2^{1\times1}}$). The complete output $\mathbf{x_{out}}$ is computed as:}
\begin{equation} 
\mathbf{x_{out}} = \Gamma\left(\left(\text{LN}(\mathbf{z'_s} + \mathbf{t_{in}}) \cdot \mathbf{W}_1^{1\times 1}\right) \cdot \mathbf{W}_1^{3\times 3}\right) \cdot \mathbf{W}_2^{1\times 1} 
\end{equation}
To further validate the argument for removing positional encodings, the impact of positional encoding was evaluated. The H-CAT encoder was replaced with a Swin Transformer network, which incorporates PE, and inference was conducted on the GLaS dataset. At a resolution of $224\times224$, the Dice Score dropped by $0.8\%$ with the use of PE, while at a higher resolution of $512\times512$, the Dice Score declined even further by $1.3\%$. These findings suggest that the proposed approach offers a more robust and effective encoding mechanism than the methods that rely on positional encoding.

\subsection{ConvNeXt Block}
The ConvNeXt blocks in the ConvNeXt encoder branch are designed to generate spatially enriched multi-scale feature maps, denoted as $\mathbf{E'_s}$, with dimensions $H/p \times W/p \times 2^{s-1} \cdot C$. As shown in Fig. \ref{figure-3}(b), these blocks use an inverted bottleneck structure, with the depth-wise convolution layer at the top to handle a large kernel size ($7 \times 7$) for fewer input channels, as suggested by \textcolor{blue}{\citep{howard2017mobilenets}}. This design facilitates the efficient use of dense $1\times1$ convolution layers to extract point-wise information along expanded channel dimensions in later stages of the architecture. While the original ResNet incorporates Batch Normalization \textcolor{blue}{\citep{ioffe2017batch}} to improve convergence and reduce overfitting, Vision Transformers (ViT) employ Layer Normalization \textcolor{blue}{\citep{ba2016layer}}, which has shown strong performance across various computer vision tasks. The claim is supported by \textcolor{blue}{\citep{liu2022convnet}}, stating that Layer Normalization (LN) enhances accuracy in ConvNeXt models. {In line with this, we also incorporate a LN block in our framework, followed by a GELU activation layer in accordance with the Vision Transformer approach.}

\begin{figure}
    \centering
    \includegraphics[width=1\linewidth]{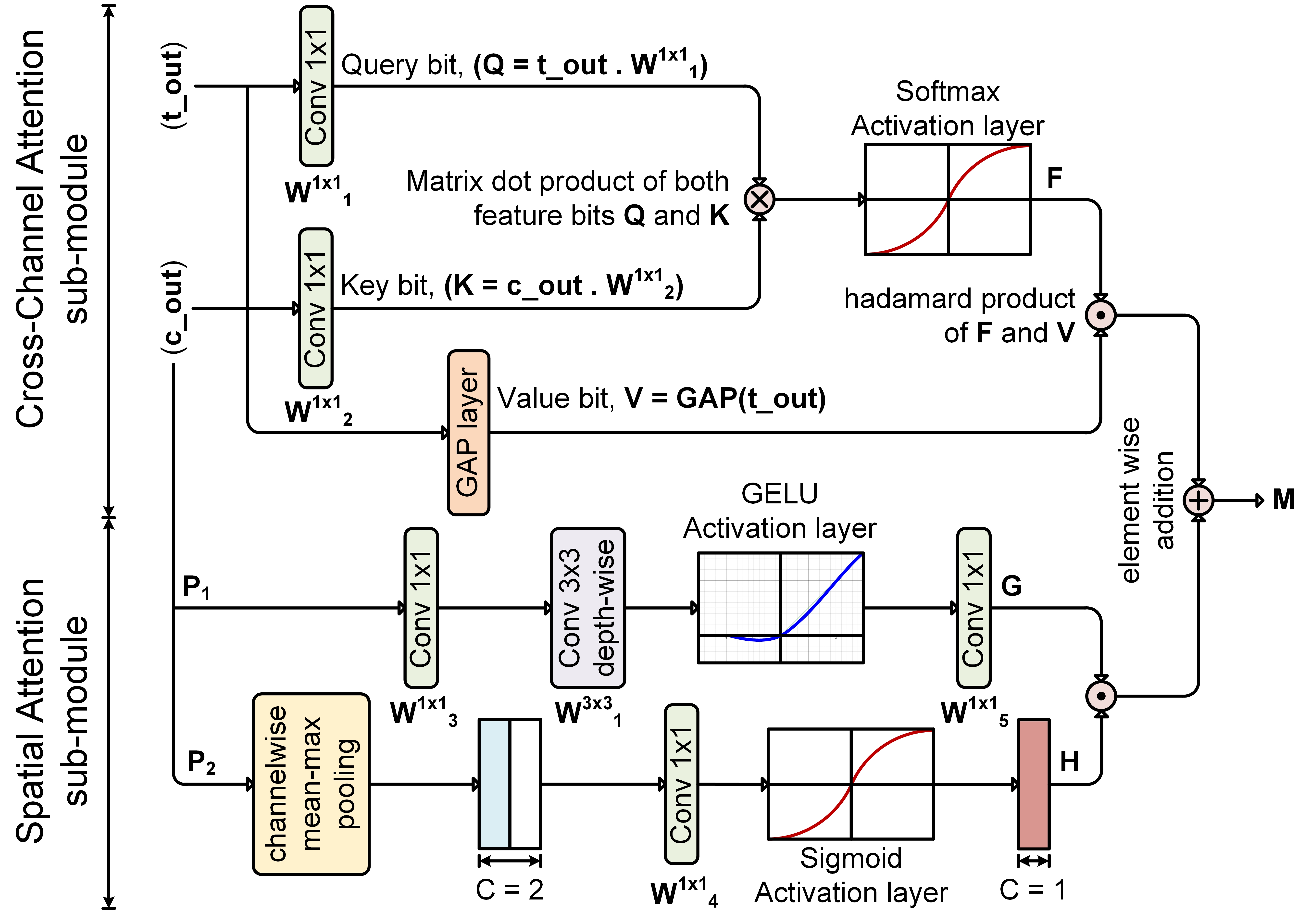}
    \caption{\textbf{Overview of the Cross Channel Trans-Convolutional Fusion Attention block.} The Cross Channel Attention module merges the outputs from the CAT and ConvNext blocks ($\mathbf{t_{out}}$ and $\mathbf{c_{out}}$) through a softmax-based fusion, enhancing global information along the channel dimension. The Spatial Attention component processes the ConvNext output through two pathways ($P_1$ and $P_2$) to remove noise and encode contextual information, preserving spatial resolution and long-range dependencies. The final output is a fusion of both channel and spatial attention, ensuring effective aggregation of multi-scale feature representations across both dimensions.}
    \label{figure-5}
\end{figure}

\subsection{Cross Channel Trans-Convolution Fusion Attention} 
A Cross Channel Trans-Convolutional Fusion Attention mechanism is proposed (see Fig. \ref{figure-5}) to integrate the intermittent feature maps produced by both the ConvNext and CAT blocks (see Section 2.1 and 2.2). This mechanism combines the two feature representations by applying Cross Channel Attention along the channel dimension and Spatial Attention along the spatial dimension. The Cross Channel Attention, resembling the scaled dot-product attention \textcolor{blue}{\citep{vaswani2017attention}}, enhances the global information embedded within the feature representation. It achieves this by merging the outputs from the CAT ($\mathbf{t_{out}}$) and ConvNext ($\mathbf{c_{out}}$) blocks using a softmax layer, as described by the following equation:
\begin{equation}
    \mathbf{F} = \mathcal{S}\left(\mathbf{t_{out}}\cdot \mathbf{W_1^{1\times 1}} \odot \mathbf{c_{out}}\cdot \mathbf{W_2^{1\times 1}}\right)
\end{equation}
This can be interpreted similarly to $\mathcal{S}(\mathbf{QK}^T)$, where $\mathbf{Q} = \mathbf{t_{out}} \cdot \mathbf{W_1^{1\times 1}}$ and $\mathbf{K} = \mathbf{c_{out}} \cdot \mathbf{W_2^{1\times 1}}$. A global average pooling operation (with a size of $1\times1\times2^{s-1}\cdot C$) is applied to $\mathbf{t_{out}}$ to define the $\mathbf{V}$ bit, which is then multiplied by $\mathbf{F}$ to maintain long-range dependencies along the channel dimension.

The Spatial Attention component ensures the preservation of spatial information along the resolution axes and reduces noise in the ConvNext feature maps ($\mathbf{c_{out}}$). To achieve this, $\mathbf{c_{out}}$ is processed through two separate pathways, $P_1$ and $P_2$. In $P_1$, a series of convolution and activation layers remove noise from the locally learned features of $\mathbf{c_{out}}$ and selectively aggregate global information to generate the output map $\mathbf{G}$. In contrast, $P_2$ encodes broader contextual positional information into the local features by producing a spatial descriptor $\mathbf{H}$ using a channel-wise pooling operation. The outputs of both pathways are multiplied and then added to the result of the Cross Channel Attention module to preserve global information across both the channel and spatial dimensions.

\subsection{Conv-G-NeXt Decoder} 
The intermittent features $\mathbf{E_s}$ generated at each encoder stage, enhanced with learned descriptors, {are transmitted through skip connections to ensure maximum information transfer to the subsequent stages in the decoder}. The primary function of the decoder stages is to accurately model localized inter-pixel dependencies in order to produce precise segmentation masks. A convolutional decoder effectively accomplishes this by utilizing upsampled latent representations $\mathbf{D_s}$, drawing local dependencies and emphasizing inherent activations. However, minimization of incorrect positive predictions for small objects with substantial shape variations remains challenging. Improvement in  accuracy is possible using the existing segmentation frameworks \textcolor{blue}{\citep{shaker2024unetr++}} by utilizing additional object localization models, splitting the task into separate localization and segmentation stages. Inspired by these approaches, the proposed decoder network is structured with two key stages: (1) a Spatial Attention Fusion Gate {(SAFG)} module and (2) Conv-G-NeXt blocks. The {SAFG} module merges upsampled and skip-connected features to promote smooth concatenation and generate spatially attended feature maps $\mathbf{D_s}$, which are then passed to the following Conv-G-NeXt block, where learnable weights are applied to the upsampled maps for the next {SAFG} module.

\begin{figure}[t]
    \centering
    \includegraphics[width=1\linewidth]{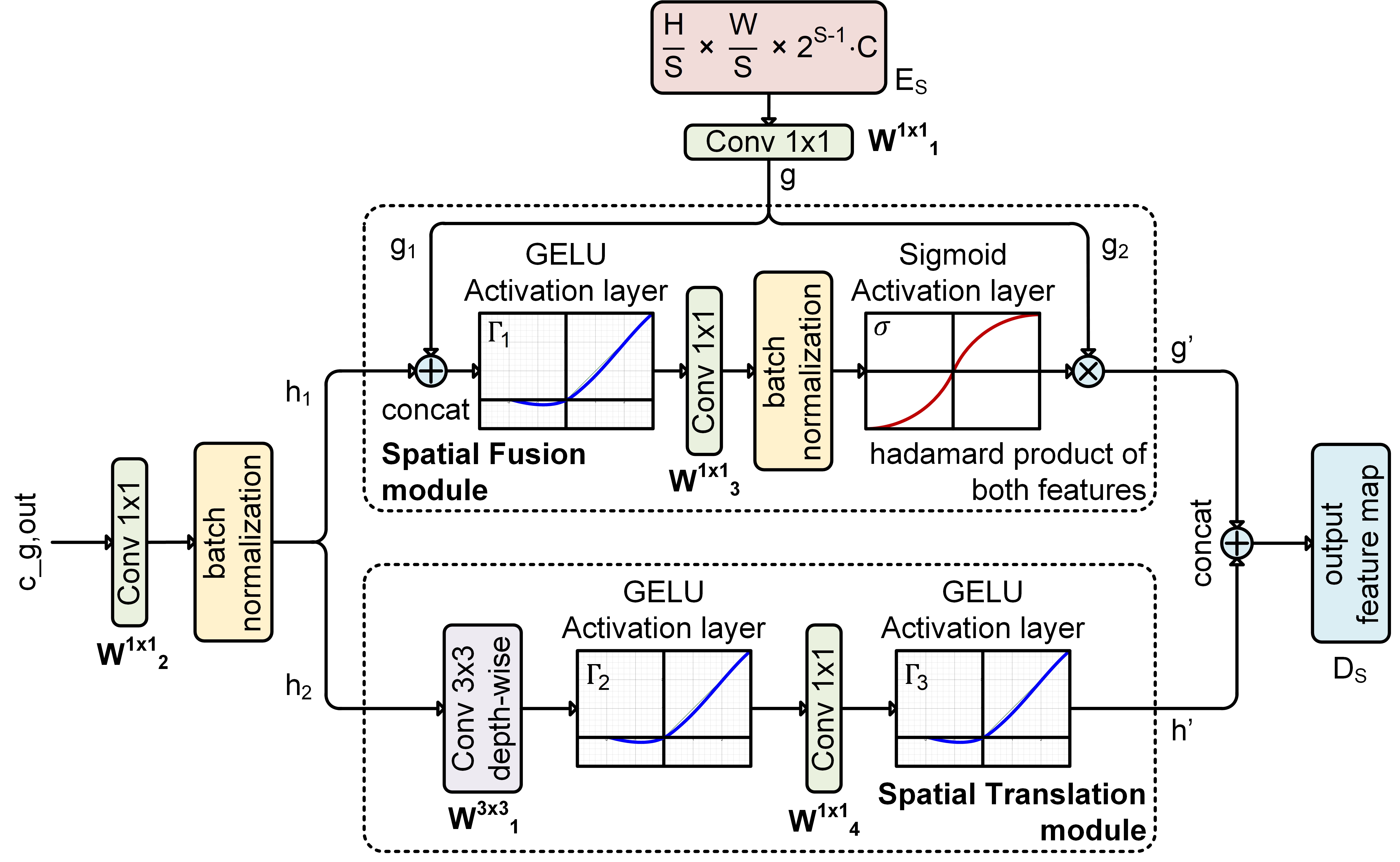}
    \caption{\textbf{Overview of the proposed Spatial Attention Fusion Gate.} The Spatial Fusion module combines the processed outputs from Conv-G-Next blocks and intermittent feature maps $\mathbf{E_s}$. The fused outputs are then passed through a convolution layer with Softmax and GeLU activations, producing output $g'$ to achieve sparser activations (Spatial Fusion module). Additionally, the Spatial Translation module ensures semantically discriminative features by processing the Conv-G-Next output through a series of depthwise and pointwise convolution layers with GeLU activation. The final output ensures rich rich learned content that prevents mis-classification.}
    \label{figure-6}
\end{figure}

\subsubsection{Spatial Attention Fusion Gate} 
Information derived from skipped maps is often contaminated with irrelevant and noisy responses. Therefore, attention gates are integrated before multi-stage CNNs to progressively suppress background noise without losing valuable information. The proposed {SAFG} (Fig. \ref{figure-6}) is embedded within the decoder architecture to enhance significant features passed through skip connections and up-sampling (refer Fig. \ref{figure-2}). {SAFG} is designed to focus on local pixels and extract relevant feature responses, identifying areas of interest. This is achieved through a gating operation that captures contextual information locally, as outlined in \textcolor{blue}{\citep{oktay2018attention, wang2017residualatt}}, performed using channel-wise $1\times 1$ convolutions with the input tensor. In image segmentation tasks \textcolor{blue}{\citep{xie2021segformer}}, sparser activations are preferred at the output. Therefore, a spatial fusion sub-module is proposed to aggregate information from $\mathbf{E_s}$ and $\mathbf{c_{g,out}}$ (the output of Conv-G-NeXt blocks) to meet this objective. Experimentally, this leads to better training convergence of the involved parameters, as shown in \textcolor{blue}{\citep{mnih2014recurrent}}.

\begin{figure}
    \centering
    \includegraphics[width=1\linewidth]{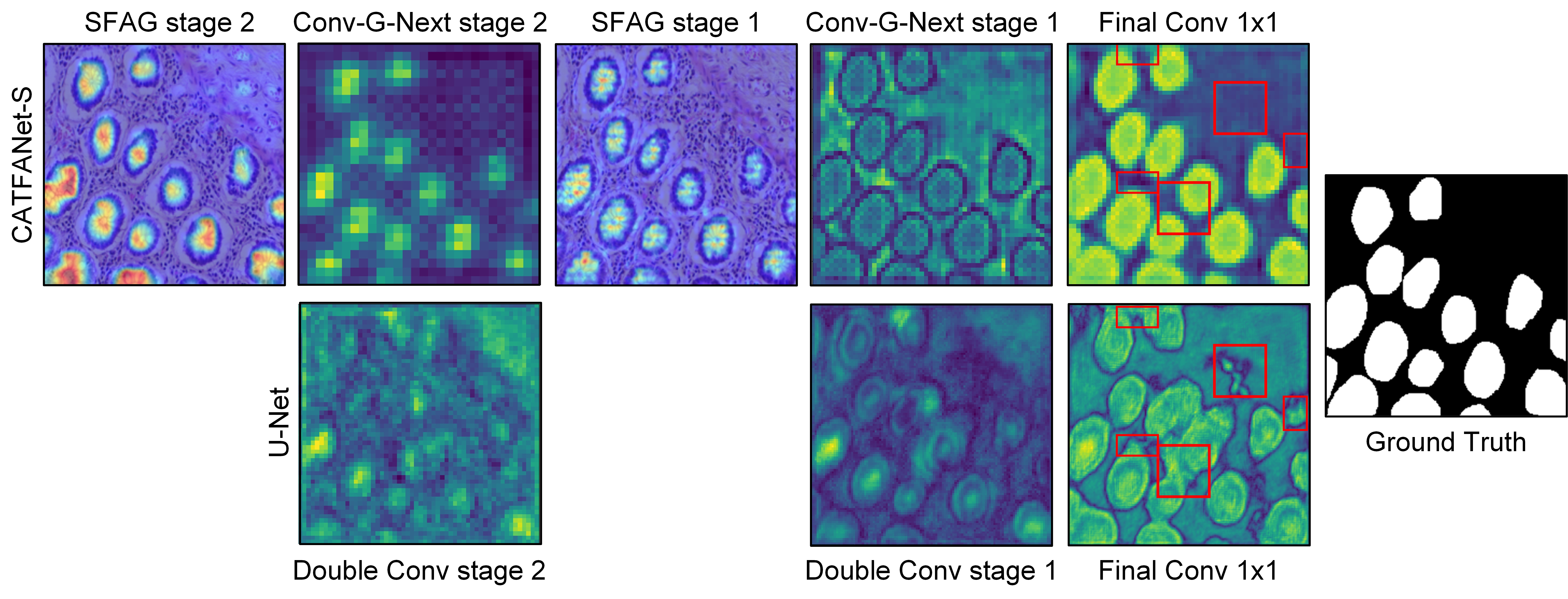}
    \caption{\textbf{Comparison of up-sampled feature maps generated by CATFA-Net-S (top) and U-Net (bottom) in the final decoder stages when evaluated using the GLaS dataset}. Red boxes indicate {major} miss-classifications in U-Net's output, which are not present in CATFA-Net-S predictions. The Spatial Fusion and Translational module in {SAFG} aids this process by reducing false predictions in upsampled feature maps.}
    \label{figure-7}
\end{figure}

As noted in \textcolor{blue}{\citep{jetley2018learn}}, low-level feature maps (such as the outputs of decoder stages) are not utilized in gating functions because they do not provide a high-dimensional representation of the input data. Consequently, we introduce a Spatial Translation Module that forces the image data to be semantically discriminative at different decoder stages. Additionally, this sub-module filters neuron activations during both forward and backward passes, enabling model parameters to be updated predominantly based on the targeted regions (see Fig. \ref{figure-7}). In summary, {SAFG} enhances the model's ability to focus on a wide range of image foreground content and prevent misclassifications.

\subsubsection{Conv-G-NeXt Block}
As previously noted, the convolutional decoder {in CATFA-Net utilizes} the proposed Conv-G-NeXt block (see Fig. \ref{figure-3}(c)), an enhanced {variant} of the ConvNeXt block \textcolor{blue}{\citep{liu2022convnet}}. {Building upon this baseline architecture, we systematically analyze and modify specific components of the original ConvNeXt design to improve decoding accuracy. The two key modifications are discussed below.}

{(1) Substituting LN with BN: While ConvNeXt originally employs Layer Normalization (LN) to align with Transformer-style architectures, Batch Normalization (BN) \textcolor{blue}{\citep{ioffe2017batch}} has long been established as effective in CNNs, accelerating convergence and reducing overfitting. To evaluate their relative impact within CATFA-Net, we conducted experiments on the GLaS dataset. Using LN in the decoder blocks yielded a Dice score of 91.7\%, whereas replacing it with BN increased the score to 92.3\%. Based on this empirical improvement, we adopt BN over LN in our Conv-G-Next block.}

{(2) Adding output non-linearity: In conventional convolutional networks, activation functions are typically appended to the outputs of convolutional layers, including $1 \times 1$ convolutions \textcolor{blue}{\citep{liu2022convnet}}. ConvNeXt, however, removes most GELU activations, retaining only one between two $1 \times 1$ convolutions (see Fig. \ref{figure-3}(b)), which results in only a modest accuracy improvement of 0.2\% \textcolor{blue}{\citep{liu2022convnet}}. In contrast, our experiments on the GLaS dataset demonstrated that reintroducing a GELU activation after the final `Conv $1 \times 1$' layer led to a more significant performance boost of 0.9\%, achieving a Dice score of 93.1\%. As a result, we include a GELU activation before collecting outputs from each decoder block in CATFA-Net.}

\vspace{-1mm}
\section{{Experiments}}
\subsection{{Evaluation Datasets}}
{We evaluate the proposed framework on six publicly available medical imaging datasets spanning a diverse range of anatomical structures and imaging modalities. These include the Gland Segmentation in Colon Histology Images (GLaS) dataset \textcolor{blue}{\citep{glasdataset}}, the Data Science Bowl (DSB) 2018 dataset \textcolor{blue}{\citep{caicedo2019nucleus}}, the CVC ClinicDB dataset \textcolor{blue}{\citep{bernal2015wm}}, the Retinal Fundus Glaucoma Challenge (REFUGE) dataset \textcolor{blue}{\citep{orlando2020refuge}}, the International Skin Imaging Collaboration (ISIC) 2018 dataset \textcolor{blue}{\citep{tschandl2018ham10000}}, and the Triple Negative Breast Cancer (TNBC) dataset \textcolor{blue}{\citep{naylor2018segmentation}}. Below, we provide a brief description of each dataset and justify their inclusion to demonstrate the generalizability of our framework across clinically relevant segmentation tasks.}

{(1) GLaS: This dataset consists of 165 images derived from 16 hematoxylin and eosin (H\&E) stained histological sections of stage T3 or T4 colorectal adenocarcinoma—the most common subtype of colorectal cancer. Each section originates from a different patient and was processed independently, introducing real-world intra- and inter-subject variability. The dataset is divided into 85 training images  and 80 test images. We follow the official benchmark protocol and retain the original train-test split. The segmentation task involves delineating glandular structures from surrounding tissue and is formulated as a binary problem. All images were resized to $224 \times 224 \times 3$ for implementation compatibility and addressing hardware constraints.}

{(2) DSB 2018: Derived from the Data Science Bowl 2018 challenge \textcolor{blue}{\citep{caicedo2019nucleus}}, this dataset includes 670 microscopy images for nuclei segmentation. The images were collected from over 16 independent experiments involving a variety of cell lines, imaging modalities (brightfield and fluorescence microscopy), staining protocols, and research centers, contributing to high visual and biological variability. We created a custom 75\%–25\% split for training (503 images) and testing (167 images). All images were resized to $224 \times 224 \times 3$ to ensure consistency with our model input dimensions.}

{(3) CVC Clinic DB: This colonoscopy video dataset comprises 612 frames extracted from 31 sequences (from 23 patients), each annotated with binary polyp segmentation masks. The dataset captures a wide range of polyp morphologies and anatomical variations, making it a valuable benchmark for colorectal lesion segmentation. We adopted a custom split of 75\% training (459 images) and 25\% testing (153 images), and resized all images to $224 \times 224 \times 3$.}

{(4) ISIC 2018: This dataset, released as part of the ISIC 2018 Challenge \textcolor{blue}{\citep{codella2019skin}}, consists of 2,596 dermoscopy images with binary segmentation masks for skin lesion boundaries. The images were acquired over two decades from two different research sites, resulting in substantial data heterogeneity. We used a 75\%–25\% split for training (1,947 images) and testing (649 images), and all images were resized to $224 \times 224 \times 3$.} 

{(5) REFUGE: Released as part of the REFUGE 2018 Challenge, this dataset includes 800 color fundus images (CFIs) acquired by ophthalmologists and technicians across several hospitals and clinical research studies. The segmentation task involves delineating the optic disc and optic cup to assess glaucomatous damage to the optic nerve head. Due to hardware constraints, the original high-resolution images were resized to $224 \times 224 \times 3$. We follow the official split of 400 images for training and 400 for test.}

{(6) TNBC: This dataset contains 50 H\&E-stained histological images of adipose tissue with high cellularity regions corresponding to invasive breast carcinoma. The samples were obtained from 11 TNBC patients at the Curie Institute. The original dataset includes dense annotations of various cell types (e.g., epithelial, myoepithelial, fibroblasts, endothelial cells). For our segmentation task, these annotations were converted into binary masks, and all images were resized to $224 \times 224 \times 3$. As this dataset is used solely for robustness evaluation, no train-test split was applied (see Section 3.6).}

{These datasets collectively provide a diverse and challenging testbed for evaluating the proposed segmentation framework, CATFA-Net against state-of-the-art models. The metrics, model architecture, and training protocol used for evaluation are described next.}

\subsection{{Evaluation Metrics, Network Structure, and Implementation Settings}}

Segmentation performance is evaluated using seven standard metrics usually adopted in medical image analysis: Dice Similarity Coefficient (DSC), Intersection over Union (IoU), Precision (P), Recall (R), Specificity (Sp), Matthews Correlation Coefficient (MCC), {and the Hausdorff Distance (HD). These metrics jointly capture both region overlap and boundary-level segmentation accuracy, enabling a comprehensive performance evaluation.} 

Two variants of the proposed CATFA-Net architecture are implemented: CATFA-Net-S and CATFA-Net-L. These variants differ primarily in the number of channels ($C$) and the allocation of CAT and ConvNeXt blocks across encoder stages (see Fig. \ref{figure-2}). Following the multi-stage design of Swin-Transformer \textcolor{blue}{\citep{liu2021swin}} and SegFormer \textcolor{blue}{\citep{xie2021segformer}}, the number of channels is doubled at each stage to enhance multiscale feature extraction and representation. The distribution of CAT and ConvNeXt blocks across stages is optimized for representational efficiency and adheres to a specific stage-wise compute ratio. The configurations are as follows:
\begin{itemize}
    \item CATFA-Net-S: $C = \{96, 192, 384, 768\}$; CAT blocks in H-CAT encoder arm follows $\{1:1:3:1\}$, ConvNeXt blocks in ConvNeXt encoder arm: $\{3:3:9:3\}$
    \item CATFA-Net-L: $C = \{128,256,512,1024\}$; CAT blocks in H-CAT encoder arm follows $\{2:2:6:2\}$, ConvNeXt blocks in ConvNeXt encoder arm: $\{3:3:27:3\}$
\end{itemize}
Both models are implemented in PyTorch and trained on a single NVIDIA Tesla V100 GPU with 32 GB of memory. The ConvNeXt encoder backbones (ConvNeXt-T for CATFA-Net-S and ConvNeXt-B for CATFA-Net-L) are initialized with pretrained weights from ImageNet-1K \textcolor{blue}{\citep{liu2022convnet}}. {For each independent run, the weights of the H-CAT encoder and Conv-G-Next decoder are randomly initialized using different seeds to assess training robustness.} Standard online data augmentation techniques, including horizontal and vertical flips and random rotations, are employed during training. Input image intensities are linearly scaled to approximate a standard normal distribution, $N(0,1)$. Training is performed with a batch size of 8. The loss function used is the generalized Dice loss \textcolor{blue}{\citep{sudre2017generalised}}, effective for handling class imbalance in binary segmentation tasks. It is computed as:
\begin{equation}
    \mathcal{L}(R, P) = 1 - \left(\frac{\sum_{n=1}^Np_nr_n + \epsilon}{\sum_{n=1}^Np_n + r_n + \epsilon}\right) - \left(\frac{\sum_{n=1}^N(1-p_n)(1-r_n) + \epsilon}{\sum_{n=1}^N2-p_n-r_n+\epsilon}\right)
\end{equation}
Here, $R$ denotes the reference segmentation with pixel values $r_n$, and $P$ represents the predicted probability map over $N$ pixels with values $p_n$, where the background probability is $1 - P$. Optimization is performed using the AdamW optimizer \textcolor{blue}{\citep{loshchilov2019decoupled}} with a learning rate of $0.0001$ over 50 epochs. {To ensure robust and meaningful performance estimates, all experiments are repeated across 10 independent runs. Each run is conducted with a different random seed to introduce variability in weight initialization and data shuffling, while keeping the dataset split fixed. Final results are reported as mean $\pm$ standard deviation, providing a reliable measure of model stability and generalization.}

\begin{figure}[ht]
    \centering
    \includegraphics[width=1.0\linewidth]{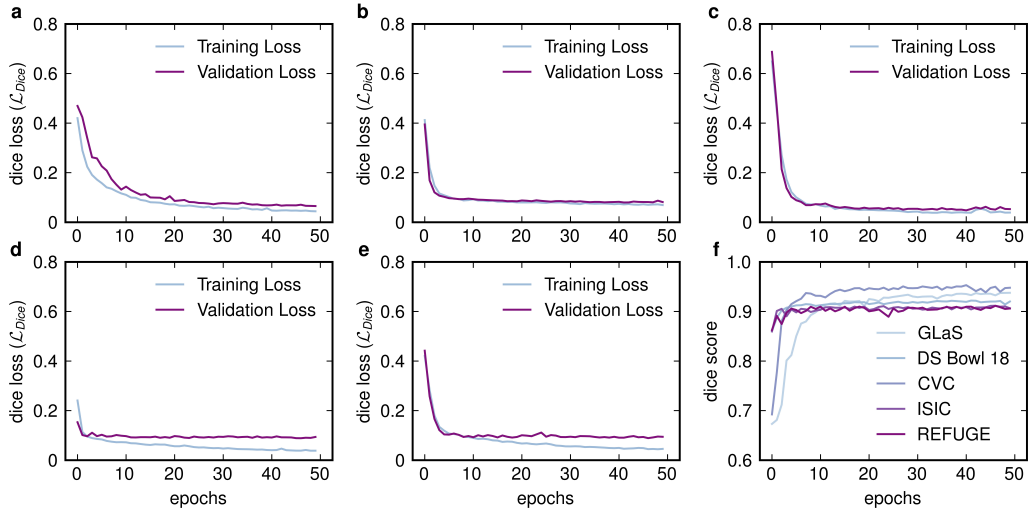}
    \caption{\textbf{Training and validation Dice loss curves for CATFA-Net-L across five publicly available datasets:} \textbf{(a)} GLaS, \textbf{(b)} DSB 2018, \textbf{(c)} CVC Clinic DB, \textbf{(d)} ISIC 2018, and \textbf{(e)} REFUGE. {\textbf{(f)} illustrates the mean Dice score convergence over 50 epochs, demonstrating the stability and robustness of CATFA-Net across diverse segmentation tasks. All curves represent the mean performance per epoch, averaged over 10 independent runs with different random seeds.}}
    \label{figure-8}
\end{figure}

\subsection{{Comparison with state-of-the-art approaches}}
{We compare CATFA-Net against twelve recent SOTA segmentation models that span convolutional, attention-enhanced, and transformer based architectures. Among convolutional baselines, U-Net \textcolor{blue}{\citep{ronneberger2015unet}} laid the foundation for encoder–decoder segmentation networks. UNet++ \textcolor{blue}{\citep{zhou2018unet++}} introduced nested skip connections for improved feature fusion, while R2U-Net \textcolor{blue}{\citep{alom2018recurrent}} leveraged recurrent residual blocks to enhance representational depth. PSPNet \textcolor{blue}{\citep{zhao2017pyramid}} and DeepLabv3+ \textcolor{blue}{\citep{chen2018encoder}} introduced pyramid-based multi-scale context modules, with the latter adding a decoder for refined boundary prediction. nnUNet \textcolor{blue}{\citep{isensee2021nnu}} provided a robust self-configuring framework that adapts architecture and training strategies to each dataset.}

{Attention-based extensions like Attention UNet \textcolor{blue}{\citep{oktay2018attention}} and Attention R2UNet \textcolor{blue}{\citep{wang2017residualatt}} incorporate attention mechanisms to focus on salient regions and enhance semantic relevance across layers. Transformer-based models mark the next stage of development. TransUNet \textcolor{blue}{\citep{chen2021transunet}} combines a ViT encoder with a CNN decoder, while SwinUNet \textcolor{blue}{\citep{cao2021swinunet}} adopts a pure transformer design with hierarchical windowed attention. Hybrid designs like UCTransNet \textcolor{blue}{\citep{UCTransNet}} and the Fully Convolutional Transformer (FCT) \textcolor{blue}{\citep{tragakis2023fullyconvolutionaltransformermedical}} aim to unify the strengths of both paradigms by integrating channel-wise transformers and hierarchical convolution-transformer fusion, respectively.}

{To assess the statistical significance of observed performance differences, we apply a one-tailed Wilcoxon signed-rank test \textcolor{blue}{\citep{he2025deep}}. This non-parametric test is appropriate for comparing paired, non-normally distributed samples and aligns with our directional hypothesis that CATFA-Net outperforms existing baselines. For each model, performance scores were collected across 10 independent runs (see section 3.2), where each run was trained using a different random seed. The best-performing checkpoint per run—based on the highest validation Dice score—was selected to construct paired samples. For fair comparison, all baseline models were implemented using official open-source code. Training and evaluation followed the dataset splits described in section 3.1 and the implementation settings in section 3.2. The resulting $p$-values, reported in Tables 1–5, are consistently below $0.01$ for most comparisons, confirming that CATFA-Net’s improvements are statistically significant and unlikely to have occurred by chance.}

\begin{table}[t]
    \centering
    \setlength{\extrarowheight}{1.0pt}
    {
    \caption{Quantitative comparison of CATFA-Net with SOTA methods on GLaS.}
    \resizebox{\textwidth}{!}{
    \begin{threeparttable}
    \begin{tabular}{lccccccc}
    \hline\hline
        Method & DSC (\%) & IoU (\%) & P (\%) & R (\%) & Sp (\%) & MCC (\%) & HD (mm) \\
    \hline
        U-Net \textcolor{blue}{\citep{ronneberger2015unet}} & $82.26 \pm 4.9^{*}$ & $72.67 \pm 6.4^{*}$ & $86.04 \pm 1.3^{*}$ & $79.49 \pm 2.9^{*}$ & $88.1 \pm 1.4^{*}$ & $66.12 \pm 5.8^{*}$ & $38.86 \pm 8.3^{*}$ \\
        U-Net++ \textcolor{blue}{\citep{zhou2018unet++}} & $88.41 \pm 5.5^{*}$ & $79.52 \pm 1.1^{\dagger}$ & $89.32 \pm 9.3^{\dagger}$ & $86.10 \pm 4.2^{*}$ & $89.85 \pm 7.1^{*}$ & $78.74 \pm 3.3^{*}$ & $34.53 \pm 3.9^{*}$ \\
        R2U-Net \textcolor{blue}{\citep{alom2018recurrent}} & $85.15 \pm 6.1^{*}$ & $75.29 \pm 3.6^{*}$ & $84.33\pm4.8^{*}$ & $87.43\pm5.9^{*}$ & $84.46 \pm 3.7^{*}$ & $72.72\pm 6.5^{*}$ & $36.15 \pm 3.2^{*}$ \\
        PSPNet \textcolor{blue}{\citep{zhao2017pyramid}} & $84.48 \pm 3.6^{*}$ & $73.54\pm4.5^{*}$ & $79.33\pm6.7^{*}$ & $89.98 \pm 7.1^{*}$ & $76.01 \pm 6.1^{*}$ & $67.42 \pm 7.7^{*}$ & $37.16 \pm 4.6^{*}$ \\
        Deeplabv3+ \textcolor{blue}{\citep{chen2018encoder}} & $83.79 \pm 7.2^{*}$ & $73.25 \pm 2.7^{\dagger}$ & $77.45 \pm 8.9^{*}$ & $90.53 \pm 9.4^{\dagger}$ & $75.62 \pm 3.3^{*}$ & $68.23 \pm 2.1^{*}$ & $37.33 \pm 2.7^{*}$ \\
        nnUNet \textcolor{blue}{\citep{isensee2021nnu}} & $89.34 \pm 4.5^{\dagger}$ & $80.71 \pm 6.1^{\ddagger}$ & $88.60 \pm 4.9^{\dagger}$ & $89.10\pm5.1^{\dagger}$ & $89.93 \pm 7.4^{\dagger}$ & $80.11\pm4.5^{\dagger}$ & $34.61\pm6.6^{*}$ \\
        AttUNet \textcolor{blue}{\citep{oktay2018attention}} & $87.02\pm1.8^{*}$ & $76.63 \pm 6.8^{*}$ & $85.16\pm5.1^{*}$ & $90.07\pm3.8^{*}$ & $81.14\pm6.9^{*}$ & $73.27 \pm 4.4^{*}$ & $35.11\pm6.9^{*}$ \\
        AttR2UNet \textcolor{blue}{\citep{wang2017residualatt}} & $87.91 \pm 3.7^{*}$ & $77.12 \pm 7.3^{*}$ & $85.21 \pm 2.6^{*}$ & $89.15 \pm 6.6^{\dagger}$ & $84.81 \pm 7.2^{*}$ & $73.42 \pm 9.2^{*}$ & $35.05 \pm 9.6^{*}$ \\
        TransUNet \textcolor{blue}{\citep{chen2021transunet}} & $90.29 \pm 2.3^{\dagger}$ & $83.24 \pm 4.2^{*}$ & $89.96 \pm 7.5^{\dagger}$ & $93.66 \pm 5.6^{\ddagger}$ & $86.33 \pm 7.8^{*}$ & $81.97 \pm 8.1^{\dagger}$ & $31.16 \pm 5.2^{\dagger}$ \\
        SwinUNet \textcolor{blue}{\citep{cao2021swinunet}} & $91.21 \pm 4.8^{\dagger}$ & $86.02 \pm 3.9^{\dagger}$ & $91.31 \pm 6.2^{\ddagger}$ & $92.25 \pm 2.3^{\dagger}$ & $92.04 \pm 5.4^{\ddagger}$ & $84.91 \pm 2.5^{\ddagger}$ & $30.28 \pm 4.3^{\dagger}$ \\
        UCTransNet \textcolor{blue}{\citep{UCTransNet}} & $93.03 \pm 2.1^{\dagger}$ & $86.01 \pm 5.0^{\ddagger}$ & $94.45 \pm 3.0^{\ddagger}$ & $93.85 \pm 4.9^{\ddagger}$ & $93.13 \pm 5.1^{\ddagger}$ & $86.39 \pm 6.3^{\dagger}$ & $30.08 \pm 2.1^{\ddagger}$ \\
        FCT \textcolor{blue}{\citep{tragakis2023fullyconvolutionaltransformermedical}} & $93.30 \pm 6.7^{\ddagger}$ & $86.30 \pm 3.2^{\ddagger}$ & $92.56 \pm 2.4^{\ddagger}$ & $94.06 \pm 6.4^{\ddagger}$ & $92.44 \pm 5.2^{\ddagger}$ & $86.22 \pm 7.1^{\ddagger}$ & $30.05 \pm 7.5^{\ddagger}$ \\
        \hline
        CATFA-Net-S & $94.04 \pm 1.9$ & $87.90 \pm 1.8$ & $94.16 \pm 3.3$ & $94.56 \pm 2.6$ & $94.6 \pm 2.5$ & $87.10 \pm 1.8$ & $29.64 \pm 3.1$ \\
        CATFA-Net-L & $\mathbf{94.48 \pm 2.4}$ & $\mathbf{89.71 \pm 2.8}$ & $\mathbf{95.30 \pm 2.2}$ & $\mathbf{95.04 \pm 2.3}$ & $\mathbf{94.68 \pm 2.1}$ & $\mathbf{88.10 \pm 2.5}$ & $\mathbf{28.20 \pm 2.5}$ \\
        \hline
    \end{tabular}
    \begin{tablenotes}
    \item $^*p$-value $<0.001$, based on one-tailed Wilcoxon signed-rank test against CATFA-Net-L
    \item $^\dagger p$-value $<0.01$, based on one-tailed Wilcoxon signed-rank test against CATFA-Net-L
    \item $^\ddagger p$-value $<0.05$, based on one-tailed Wilcoxon signed-rank test against CATFA-Net-L
    \end{tablenotes}
    \end{threeparttable}}}
    \label{table-1}
\end{table}

\begin{figure}[t]
    \centering
    \includegraphics[width=1.0\linewidth]{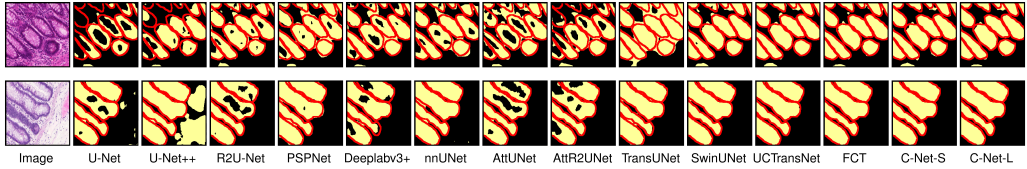}
    \caption{{\textbf{Qualitative comparison of CATFA-Net (C-Net-S and C-Net-L) variants with other SOTA methods on the GLaS Dataset.} The red curve represents the ground-truth boundary overlaid on the segmented mask produced by each model.}}
    \label{figure-9}
\end{figure}

\subsubsection{{Performance on the GLaS Dataset}}
{As shown in Table 1, CATFA-Net variants achieve top performance across all evaluation metrics on the GLaS dataset. CATFA-Net-L attains the highest mean Dice (94.48\%), IoU (89.71\%), and MCC (88.10\%), along with the lowest mean Hausdorff Distance (28.20 mm), indicating both precise region overlap and accurate boundary delineation. Compared to recent trans-convolutional hybrids such as UCTransNet and FCT, CATFA-Net-L offers measurable improvements—exceeding FCT by 1.18\% in Dice, 3.41\% in IoU, and reducing HD by 1.85 mm. These gains, accompanied by lower standard deviations, reflect the robustness and consistency of the architecture. Statistical significance is confirmed via one-tailed Wilcoxon signed-rank testing, with $p$-values below 0.01 in nearly all comparisons, except for nnUNet, validating that the improvements are unlikely to arise from random variation. To further analyze model behavior, Fig. \ref{figure-8}(a) plots the training and validation Dice loss curves over 50 epochs. A small yet stable gap is observed, suggesting minor generalization disparity that does not increase with training. The high and consistent Dice convergence (Fig. \ref{figure-8}(f)) implies that underfitting is unlikely. This gap may instead result from annotation noise or inter-sample variability. Indeed, the GLaS dataset contains highly irregular glandular morphologies and a limited number of samples, both of which are known to limit generalization on the validation set.}

{Fig. \ref{figure-9} illustrates qualitative segmentation results across models. Earlier convolutional baselines often produce incomplete or jagged gland boundaries, especially in cluttered tissue regions. While attention- and transformer-based models show improvements in continuity and suppression of background artifacts, many still exhibit minor leakages or irregular shapes. In contrast, both CATFA-Net-S and CATFA-Net-L yield cleaner and more structurally faithful segmentations, with contours closely matching ground truth across diverse samples. These visual trends reinforce the quantitative findings and underscore CATFA-Net’s effectiveness in modeling complex gland morphology with high fidelity and reliability.}

\begin{table}
    \centering
    \setlength{\extrarowheight}{1.0pt}
    {
    \caption{Quantitative comparison of CATFA-Net with SOTA methods on DSB 2018.}
    \resizebox{\textwidth}{!}{
    \begin{threeparttable}
    \begin{tabular}{lccccccc}
    \hline\hline
        Method & DSC (\%) & IoU (\%) & P (\%) & R (\%) & Sp (\%) & MCC (\%) & HD (mm) \\
    \hline
        U-Net \textcolor{blue}{\citep{ronneberger2015unet}} & $87.29 \pm 3.1^{*}$ & $79.66 \pm 4.8^{*}$ & $90.84 \pm 7.4^{*}$ & $83.51 \pm 6.4^{*}$ & $99.08 \pm 2.6^{*}$ & $87.16 \pm 5.1^{*}$ & $25.51 \pm 3.4^{*}$ \\
        U-Net++ \textcolor{blue}{\citep{zhou2018unet++}} & $90.87 \pm 3.9^{\dagger}$ & $83.31 \pm 5.9^{*}$ & $92.05 \pm 2.9^{\ddagger }$ & $89.74 \pm 3.9^{*}$ & $98.48 \pm 2.7^{*}$ & $87.2 \pm 4.3^{*}$ & $23.43 \pm 4.7^{*}$ \\
        R2U-Net \textcolor{blue}{\citep{alom2018recurrent}} & $90.7 \pm 2.4^{*}$ & $83.29 \pm 3.5^{*}$ & $91.81\pm4.9^{*}$ & $89.71\pm2.2^{*}$ & $98.42 \pm 3.4^{*}$ & $89.24\pm 2.5^{*}$ & $25.15 \pm 4.1^{*}$ \\
        PSPNet \textcolor{blue}{\citep{zhao2017pyramid}} & $87.58 \pm 5.3^{*}$ & $75.96\pm4.1^{*}$ & $85.71\pm3.4^{*}$ & $86.88 \pm 6.1^{*}$ & $97.24 \pm 3.1^{*}$ & $83.72 \pm 5.9^{*}$ & $26.24 \pm 1.7^{*}$ \\
        Deeplabv3+ \textcolor{blue}{\citep{chen2018encoder}} & $88.67 \pm 6.2^{*}$ & $80.72 \pm 6.3^{*}$ & $90.66 \pm 2.7^{\dagger}$ & $87.91 \pm 4.7^{*}$ & $98.34 \pm 4.2^{\ddagger}$ & $87.39 \pm 3.6^{\dagger}$ & $26.78 \pm 3.6^{*}$ \\
        nnUNet \textcolor{blue}{\citep{isensee2021nnu}} & $90.87 \pm 4.1^{\dagger}$ & $83.1 \pm 4.6^{\dagger}$ & $90.5 \pm 4.1^{\ddagger}$ & $90.14\pm5^{*}$ & $97.88 \pm 3.1^{\dagger}$ & $88.6\pm6.5^{\dagger}$ & $22.11\pm5^{\ddagger}$ \\
        AttUNet \textcolor{blue}{\citep{oktay2018attention}} & $91.75\pm3.7^{\ddagger}$ & $85.09 \pm 2.2^{\dagger}$ & $92.09\pm4.2^{*}$ & $91.57\pm4.1^{*}$ & $98.47\pm3.8^{\dagger}$ & $90.44 \pm 1.7^{\ddagger}$ & $24.25\pm6.9^{*}$ \\
        AttR2UNet \textcolor{blue}{\citep{wang2017residualatt}} & $91.01 \pm 4.6^{*}$ & $83.58 \pm 2.8^{\dagger}$ & $92.23 \pm 3.8^{\ddagger}$ & $89.98 \pm 2.4^{\dagger}$ & $98.62 \pm 2.3^{*}$ & $89.56 \pm 6.3^{*}$ & $22.65 \pm 3.5^{\dagger}$ \\
        TransUNet \textcolor{blue}{\citep{chen2021transunet}} & $91.23 \pm 6.3^{\dagger}$ & $84.49 \pm 3.0^{\dagger}$ & $91.01 \pm 2.2^{*}$ & $91.66 \pm 3.8^{\dagger}$ & $97.36 \pm 5^{\ddagger}$ & $89.13 \pm 3.1^{*}$ & $20.86 \pm 5.9^{*}$ \\
        SwinUNet \textcolor{blue}{\citep{cao2021swinunet}} & $92.07 \pm 2.8^{*}$ & $85.26 \pm 4.5^{\dagger}$ & $92.21 \pm 6.1^{*}$ & $91.74 \pm 6^{\dagger}$ & $98.45 \pm 3.6^{\ddagger}$ & $90.56 \pm 2.1^{\ddagger}$ & $19.58 \pm 2^{\dagger}$ \\
        UCTransNet \textcolor{blue}{\citep{UCTransNet}} & $92.23 \pm 5.7^{\ddagger}$ & $85.29 \pm 1.9^{\dagger}$ & $92.66 \pm 4^{\dagger}$ & $91.84 \pm 5.4^{\ddagger}$ & $98.49 \pm 1.8^{\ddagger}$ & $90.73 \pm 6.8^{\dagger}$ & $19.35 \pm 1.6^{*}$ \\
        FCT \textcolor{blue}{\citep{tragakis2023fullyconvolutionaltransformermedical}} & $91.04 \pm 6.5^{\dagger}$ & $84.04 \pm 3.3^{\ddagger}$ & $91.32 \pm 5.8^{\ddagger}$ & $91.56 \pm 7.2^{\dagger}$ & $97.21 \pm 4.1^{\dagger}$ & $89.14 \pm 6.6^{\ddagger}$ & $17.44 \pm 6.8^{\ddagger}$ \\
        \hline
        CATFA-Net-S & $92.03 \pm 3.1$ & $85.27 \pm 2.3$ & $\mathbf{92.74 \pm 2.7}$ & $\mathbf{93.11 \pm 2}$ & $98.23 \pm 1.5$ & $90.51 \pm 3.2$ & $18.85 \pm 1.9$ \\
        CATFA-Net-L & $\mathbf{92.14 \pm 2.6}$ & $\mathbf{85.51 \pm 2.1}$ & $92.12 \pm 3.2$ & $92.27 \pm 3.3$ & $\mathbf{98.51 \pm 1.5}$ & $\mathbf{90.64 \pm 2.9}$ & $\mathbf{17.55 \pm 3}$ \\
        \hline
    \end{tabular}
    \begin{tablenotes}
    \item $^*p$-value $<0.001$, based on one-tailed Wilcoxon signed-rank test against CATFA-Net-L
    \item $^\dagger p$-value $<0.01$, based on one-tailed Wilcoxon signed-rank test against CATFA-Net-L
    \item $^\ddagger p$-value $<0.05$, based on one-tailed Wilcoxon signed-rank test against CATFA-Net-L
    \end{tablenotes}
    \end{threeparttable}}}
    \label{table-2}
\end{table}

\begin{figure}[t]
    \centering
    \includegraphics[width=1.0\linewidth]{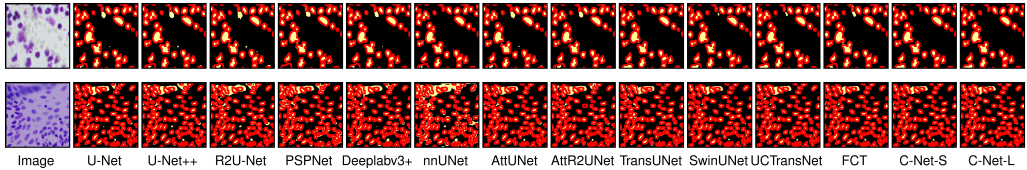}
    \caption{{\textbf{Qualitative comparison of CATFA-Net (C-Net-S and C-Net-L) variants with other SOTA methods on the DSB 2018 Dataset.} The red curve represents the ground-truth boundary overlaid on the segmented mask produced by each model.}}
    \label{figure-10}
\end{figure}

\subsubsection{Performance on the DSB 2018 dataset}
{Table 2 reports the performance of all evaluated models on the DSB 2018 dataset. CATFA-Net-L achieves the best results in five of the seven metrics, including Dice (92.14\%), IoU (85.51\%), Specificity (98.51\%), MCC (90.64\%), and Hausdorff Distance (17.55 mm). CATFA-Net-S records the highest Precision (92.74\%) and Recall (93.11\%), highlighting its ability to localize and segment nuclei with high sensitivity. Both variants outperform recent transformer hybrids such as FCT and UCTransNet, with CATFA-Net-L achieving a 0.91\% gain in Dice and a 0.49\% reduction in HD over FCT. Standard deviations are consistently low across all metrics, suggesting stable performance across diverse microscopy samples. These quantitative gains are statistically validated via the one-tailed Wilcoxon signed-rank test, with $p$-values below 0.01 for all major comparisons, confirming the significance of the observed improvements. Fig. \ref{figure-8}(b) further illustrates the training dynamics: both training and validation Dice loss curves for the DSB dataset converge smoothly, showing minimal divergence and indicating good generalization without signs of overfitting. This observation is reinforced by the Dice score trajectories in Fig. \ref{figure-8}(f), which plateau early and remain consistent across epochs.}

{Fig. \ref{figure-10} presents qualitative segmentation results on representative samples from the DSB 2018 dataset. Traditional convolutional models such as U-Net and PSPNet exhibit under-segmentation and frequent merging of adjacent nuclei, particularly in densely packed regions. While attention-enhanced and transformer-based models (e.g., TransUNet, SwinUNet) demonstrate improved boundary localization and separation, occasional shape distortions and segmentation noise are still observed. In contrast, CATFA-Net produces consistently well-contoured and well-separated nuclear boundaries that visually align closely with the ground truth, reflecting strong boundary sensitivity and morphological consistency. However, a notable failure case is observed in the second-row example, where a shadow artifact in the top-left region partially obscures several nuclei. All models, including CATFA-Net, fail to recognize this as an anomaly, misinterpreting the shaded region as part of the nuclei region. This highlights the limitations of current models when dealing with subtle acquisition-related artifacts.}

\begin{table}
    \centering
    \setlength{\extrarowheight}{1.0pt}
    {
    \caption{Quantitative comparison of CATFA-Net with SOTA methods on CVC Clinic DB.}
    \resizebox{\textwidth}{!}{
    \begin{threeparttable}
    \begin{tabular}{lccccccc}
    \hline\hline
        Method & DSC (\%) & IoU (\%) & P (\%) & R (\%) & Sp (\%) & MCC (\%) & HD (mm) \\
    \hline
        U-Net \textcolor{blue}{\citep{ronneberger2015unet}} & $86.59 \pm 3.1^{*}$ & $76.66 \pm 4.2^{*}$ & $83.75 \pm 2.5^{*}$ & $88.47 \pm 3.4^{*}$ & $97.19 \pm 2.8^{*}$ & $86.36 \pm 4.7^{*}$ & $20.25 \pm 2.7^{*}$ \\
        U-Net++ \textcolor{blue}{\citep{zhou2018unet++}} & $90.75 \pm 5.7^{*}$ & $81.13 \pm 3.8^{*}$ & $89.15 \pm 2.8^{*}$ & $88.01 \pm 3.2^{*}$ & $98.62 \pm 5.1^{*}$ & $87.13 \pm 2^{*}$ & $16.75 \pm 3.8^{*}$ \\
        R2U-Net \textcolor{blue}{\citep{alom2018recurrent}} & $87.03 \pm 2.3^{*}$ & $78.58 \pm 2.8^{*}$ & $86.18\pm2.4^{*}$ & $88.37\pm3.5^{*}$ & $98.20 \pm 3.2^{*}$ & $87.58\pm 3.2^{*}$ & $18.65 \pm 3.1^{*}$ \\
        PSPNet \textcolor{blue}{\citep{zhao2017pyramid}} & $88.17 \pm 5.3^{*}$ & $79.41\pm3.6^{*}$ & $90.42\pm5.2^{*}$ & $86.35 \pm 2.9^{*}$ & $98.31 \pm 3.7^{*}$ & $87.42 \pm 4.6^{*}$ & $19.24 \pm 2.4^{*}$ \\
        Deeplabv3+ \textcolor{blue}{\citep{chen2018encoder}} & $86.35 \pm 2.5^{*}$ & $78.61 \pm 2.7^{*}$ & $87.58 \pm 4.5^{*}$ & $85.12 \pm 4.9^{*}$ & $98.21 \pm 1.8^{*}$ & $86.19 \pm 2.2^{*}$ & $20.48 \pm 4^{*}$ \\
        nnUNet \textcolor{blue}{\citep{isensee2021nnu}} & $91.23 \pm 6.3^{\dagger}$ & $83.51 \pm 2.6^{*}$ & $90.44 \pm 6.4^{\ddagger}$ & $87.77\pm7.1^{*}$ & $98.55 \pm 2.9^{\dagger}$ & $89.22\pm4.7^{\ddagger}$ & $16.17\pm4.4^{*}$ \\
        AttUNet \textcolor{blue}{\citep{oktay2018attention}} & $88.15\pm3.9^{*}$ & $78.78 \pm 1.9^{*}$ & $90.34\pm3.7^{*}$ & $86.42\pm4.6^{*}$ & $98.55\pm2.5^{\dagger}$ & $87.62 \pm 3^{*}$ & $18.95\pm4.3^{*}$ \\
        AttR2UNet \textcolor{blue}{\citep{wang2017residualatt}} & $88.54 \pm 4.2^{*}$ & $79.11 \pm 4.9^{*}$ & $90.67 \pm 5.7^{*}$ & $85.32 \pm 2.7^{*}$ & $98.63 \pm 4.1^{\ddagger}$ & $88.12 \pm 5.1^{\dagger}$ & $19.29 \pm 3.9^{*}$ \\
        TransUNet \textcolor{blue}{\citep{chen2021transunet}} & $92.28 \pm 4.1^{*}$ & $87.65 \pm 5.6^{*}$ & $93.25 \pm 4.1^{\dagger}$ & $94.25 \pm 5.3^{\ddagger}$ & $99.19 \pm 2.3^{\ddagger}$ & $92.78 \pm 4.9^{\dagger}$ & $12.66 \pm 4.4^{*}$ \\
        SwinUNet \textcolor{blue}{\citep{cao2021swinunet}} & $93.88 \pm 2.8^{*}$ & $88.56 \pm 4.4^{\dagger}$ & $94.45 \pm 2.7^{*}$ & $93.08 \pm 3.9^{*}$ & $\mathbf{99.65 \pm 2}$ & $93.64 \pm 2.7^{\dagger}$ & $12.58 \pm 3.6^{\ddagger}$ \\
        UCTransNet \textcolor{blue}{\citep{UCTransNet}} & $94.17 \pm 3.8^{\dagger}$ & $89.46 \pm 2.3^{\dagger}$ & $94.57 \pm 5.8^{*}$ & $93.53 \pm 1.9^{\ddagger}$ & $99.25 \pm 3.3^{\ddagger}$ & $93.18 \pm 3.3^{*}$ & $13.28 \pm 2.7^{\dagger}$ \\
        FCT \textcolor{blue}{\citep{tragakis2023fullyconvolutionaltransformermedical}} & $91.04 \pm 6.5^{\dagger}$ & $84.04 \pm 3.3^{\dagger}$ & $91.32 \pm 5.8^{\ddagger}$ & $91.56 \pm 7.2^{\dagger}$ & $97.21 \pm 4.1^{\ddagger}$ & $89.14 \pm 6.6^{\dagger}$ & $17.44 \pm 6.8^{\dagger}$ \\
        \hline
        CATFA-Net-S & $95.07 \pm 3.0$ & $90.61 \pm 2.6$ & $\mathbf{95.17 \pm 2.2}$ & $94.98 \pm 2.4$ & $99.33 \pm 2.9$ & $94.65 \pm 2.5$ & $\mathbf{11.48 \pm 2.9}$ \\
        CATFA-Net-L & $\mathbf{95.3 \pm 2.3}$ & $\mathbf{91.04 \pm 2.7}$ & $94.41 \pm 3.1$ & $\mathbf{96.24 \pm 2.6}$ & $99.5 \pm 2.7$ & $\mathbf{94.89 \pm 3}$ & $12.11 \pm 1.1$ \\
        \hline
    \end{tabular}
    \begin{tablenotes}
    \item $^*p$-value $<0.001$, based on one-tailed Wilcoxon signed-rank test against CATFA-Net-L
    \item $^\dagger p$-value $<0.01$, based on one-tailed Wilcoxon signed-rank test against CATFA-Net-L
    \item $^\ddagger p$-value $<0.05$, based on one-tailed Wilcoxon signed-rank test against CATFA-Net-L
    \end{tablenotes}
    \end{threeparttable}}}
    \label{table-3}
\end{table}

\begin{figure}[t]
    \centering
    \includegraphics[width=1.0\linewidth]{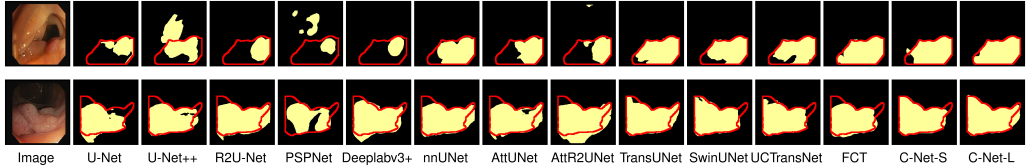}
    \caption{{\textbf{Qualitative comparison of CATFA-Net (C-Net-S and C-Net-L) variants with other SOTA methods on the CVC Clinic DB Dataset.} The red curve represents the ground-truth boundary overlaid on the segmented mask produced by each model.}}
    \label{figure-11}
\end{figure}

\subsubsection{Performance on the CVC Clinic DB dataset}
{Table 3 presents the performance comparison of CATFA-Net with SOTA segmentation models on the CVC Clinic DB dataset. CATFA-Net-L achieves top performance across four reported metrics, including a mean Dice score of 95.3\%, IoU of 91.04\%, MCC of 94.89\%, and Recall of 96.24\%. The architecture also yields a low Hausdorff distance of 12.11 mm, reflecting accurate contour alignment with ground truth. However, CATFA-Net-S establishes the best HD performance of 11.48 mm, effectively underscoring the efficacy of the hybrid design across variants. Both models outperform convolutional baselines such as U-Net++ and R2U-Net, attention-based models like AttUNet, and recent transformer-integrated designs including UCTransNet and FCT. Notably, CATFA-Net-L delivers a 1.13\% Dice improvement and a 1.71\% MCC increase over UCTransNet, while offering finer boundary precision (reduced HD by over 1 mm). All observed gains are statistically significant, with $p$-values consistently $< 0.01$ based on Wilcoxon signed-rank testing. The training and validation Dice loss curves in Fig. \ref{figure-8}(c) demonstrate rapid and stable convergence. The gap between training and validation loss remains narrow and does not exhibit divergence over time, further reinforcing the model's generalization capabilities. Final Dice scores plateau above 0.95, showing minimal fluctuation across epochs and seeds (Fig. \ref{figure-8}(f)).}

{Figure \ref{figure-11} provides a qualitative comparison across representative samples. Classical convolutional algorithms exhibit segmentation errors including false negatives and inaccurate polyp boundaries. While attention- and transformer-enhanced models improve structural consistency, some suffer from coarse edges or over-smoothing. In contrast, CATFA-Net predictions capture fine boundary contours with high precision, maintaining structural integrity even under occlusion and low-contrast conditions. For instance, in the first-row sample, CATFA-Net-L shows minimal boundary leakage and provides smooth contour delineation compared to the irregular outputs of most baselines. These qualitative results corroborate the superior boundary-aware performance evident in HD metrics and reinforce CATFA-Net’s ability to model complex, clinically relevant morphology under variable imaging conditions.}

\begin{table}
    \centering
    \setlength{\extrarowheight}{1.0pt}
    {
    \caption{Quantitative comparison of CATFA-Net with SOTA methods on ISIC 2018.}
    \resizebox{\textwidth}{!}{
    \begin{threeparttable}
    \begin{tabular}{lccccccc}
    \hline\hline
        Method & DSC (\%) & IoU (\%) & P (\%) & R (\%) & Sp (\%) & MCC (\%) & HD (mm) \\
    \hline
        U-Net \textcolor{blue}{\citep{ronneberger2015unet}} & $87.12 \pm 4.7^{*}$ & $78.25 \pm 6.2^{*}$ & $90.56 \pm 3.6^{*}$ & $83.55 \pm 4.1^{*}$ & $97.11 \pm 5.5^{*}$ & $84.62 \pm 6.8^{*}$ & $24.35 \pm 3.9^{*}$ \\
        U-Net++ \textcolor{blue}{\citep{zhou2018unet++}} & $89.12 \pm 4.2^{\dagger}$ & $80.54 \pm 8.9^{*}$ & $89.53 \pm 6.3^{\dagger}$ & $88.48 \pm 4.9^{*}$ & $96.6 \pm 4^{*}$ & $86.46 \pm 6^{*}$ & $19.1 \pm 9.1^{*}$ \\
        R2U-Net \textcolor{blue}{\citep{alom2018recurrent}} & $89.61 \pm 8.5^{*}$ & $80.46 \pm4.8^{*}$ & $92.14\pm7.7^{*}$ & $86.03\pm8.8^{\dagger}$ & $98.24 \pm1.9^{*}$ & $86.93\pm 7.1^{*}$ & $23.65 \pm 2.7^{\dagger}$ \\
        PSPNet \textcolor{blue}{\citep{zhao2017pyramid}} & $87.26 \pm 2.3^{*}$ & $79.29\pm3.5^{\dagger}$ & $89.18\pm2.2^{\ddagger}$ & $86.31 \pm 7.6^{*}$ & $97.44 \pm 3.3^{\dagger}$ & $85.66 \pm 3.9^{*}$ & $23.05 \pm 4.1^{*}$ \\
        Deeplabv3+ \textcolor{blue}{\citep{chen2018encoder}} & $87.94 \pm 6.9^{*}$ & $81.1 \pm 2.7^{*}$ & $91.22 \pm 9.4^{\dagger}$ & $86.3 \pm 1.9^{*}$ & $97.19 \pm 6.1^{\ddagger}$ & $87.17 \pm 6.3^{\dagger}$ & $24.43 \pm 4.4^{*}$ \\
        nnUNet \textcolor{blue}{\citep{isensee2021nnu}} & $89.56 \pm 3.8^{*}$ & $80.26 \pm 6.3^{\dagger}$ & $88.85 \pm 6.7^{\ddagger}$ & $86.44\pm3.6^{\dagger}$ & $96.44 \pm 8.1^{\dagger}$ & $86.17\pm5.5^{*}$ & $20.61\pm6.4^{*}$ \\
        AttUNet \textcolor{blue}{\citep{oktay2018attention}} & $87.97\pm6.1^{*}$ & $79.31\pm5.1^{*}$ & $89.66\pm4^{\dagger}$ & $88.91\pm5.3^{\ddagger}$ & $96.81\pm2.8^{*}$ & $85.53\pm5.1^{\ddagger}$ & $21.5\pm4.6^{\dagger}$ \\
        AttR2UNet \textcolor{blue}{\citep{wang2017residualatt}} & $88.86\pm1.8^{*}$ & $81.16\pm6.4^{\dagger}$ & $89.22\pm1.5^{\ddagger}$ & $89.1\pm6.7^{\dagger}$ & $97.53\pm 5.2^{\dagger}$ & $86.33\pm3.7^{*}$ & $19.45\pm5.5^{*}$ \\
        TransUNet \textcolor{blue}{\citep{chen2021transunet}} & $90.64 \pm 9.1^{*}$ & $81.66\pm2.4^{\dagger}$ & $91.15\pm3.1^{*}$ & $89.16\pm1.3^{\ddagger}$ & $97.56\pm2.6^{\dagger}$ & $87\pm 4.6^{*}$ & $16.56\pm3.3^{\dagger}$ \\
        SwinUNet \textcolor{blue}{\citep{cao2021swinunet}} & $90.54 \pm 7.3^{\dagger}$ & $83.53\pm7.5^{\dagger}$ & $91.42\pm5.7^{\ddagger}$ & $90.3\pm9.2^{\dagger}$ & $97.33\pm 2.2^{\ddagger}$ & $88.04\pm3^{*}$ & $16.29\pm3.2^{\ddagger}$ \\
        UCTransNet \textcolor{blue}{\citep{UCTransNet}} & $90.22\pm 1.9^{\dagger}$ & $83.1\pm3.3^{\dagger}$ & $\mathbf{92.91\pm 2.9}$ & $89.16\pm2.8^{\ddagger}$ & $\mathbf{98.33\pm1.8}$ & $\mathbf{88.39\pm5.1}$ & $15.21\pm2.8^{\ddagger}$ \\
        FCT \textcolor{blue}{\citep{tragakis2023fullyconvolutionaltransformermedical}} & $90.11\pm3.8^{*}$ & $82.78\pm2.1^{\dagger}$ & $91.4\pm7.2^{\ddagger}$ & $89.33\pm7.9^{\ddagger}$ & $97.14\pm 1.3^{\dagger}$ & $88.15\pm2.2^{\ddagger}$ & $15.22\pm6.6^{\dagger}$ \\
        \hline
        CATFA-Net-S & $91.27 \pm 2.6$ & $83.56 \pm 2.9$ & $92.14 \pm 2.7$ & $90.36 \pm 2.8$ & $97.22 \pm 1.7$ & $88.66 \pm 2.5$ & $14.66 \pm 4.4$ \\
        CATFA-Net-L & $\mathbf{91.55 \pm 2.2}$ & $\mathbf{83.85 \pm 1.7}$ & $92.87 \pm 3.3$ & $\mathbf{90.56 \pm 2.5}$ & $97.16 \pm 2.9$ & $\mathbf{89.55 \pm 3.2}$ & $\mathbf{13.53 \pm 2.1}$ \\
        \hline
    \end{tabular}
    \begin{tablenotes}
    \item $^*p$-value $<0.001$, based on one-tailed Wilcoxon signed-rank test against CATFA-Net-L
    \item $^\dagger p$-value $<0.01$, based on one-tailed Wilcoxon signed-rank test against CATFA-Net-L
    \item $^\ddagger p$-value $<0.05$, based on one-tailed Wilcoxon signed-rank test against CATFA-Net-L
    \end{tablenotes}
    \end{threeparttable}}}
    \label{table-4}
\end{table}

\begin{figure}[t]
    \centering
    \includegraphics[width=1.0\linewidth]{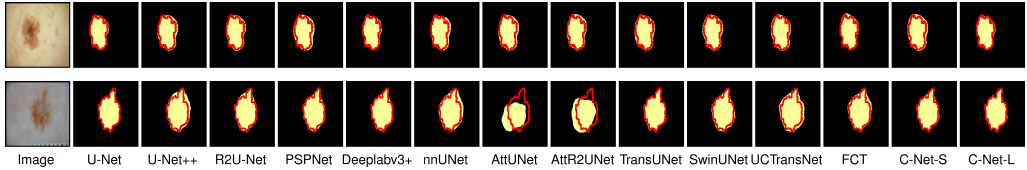}
    \caption{{\textbf{Qualitative comparison of CATFA-Net (C-Net-S and C-Net-L) variants with other SOTA methods on the ISIC 2018 Dataset.} The red curve represents the ground-truth boundary overlaid on the segmented mask produced by each model.}}
    \label{figure-12}
\end{figure}

\subsubsection{Performance on the ISIC 2018 dataset}
{As presented in Table 4, CATFA-Net exhibits strong performance on the ISIC 2018 skin lesion segmentation task. CATFA-Net-L achieves the highest mean Dice score (91.55\%) and IoU (83.85\%), along with the lowest Hausdorff Distance (13.53 mm), outperforming all twelve baseline models across key metrics. It also records the highest MCC (89.55\%), suggesting strong agreement between predicted and ground truth segmentations. While UCTransNet and SwinUNet approach CATFA-Net’s performance, CATFA-Net-L achieves a better trade-off between region-level accuracy, boundary alignment, and robustness—reflected in its consistently low standard deviations. Fig. \ref{figure-8}(d) illustrates the training and validation Dice loss curves for ISIC 2018. Both losses show mild divergence from each other but remain tightly aligned over the course of training, indicating mild overfitting. This behavior can be attributed to class imbalance and irregular morphologies in the ISIC 2018 Dataset.}

{Fig. \ref{figure-12} presents a qualitative comparison across representative samples. Traditional CNN-based architectures like show imprecise boundary predictions and segmentation leaks into surrounding tissue. Transformer-integrated models such as SwinUNet and FCT improve lesion contour sharpness but occasionally suffer from over-segmentation. In contrast, CATFA-Net-L consistently generates well-aligned, smooth lesion boundaries closely following the ground truth. These results further confirm CATFA-Net’s ability to handle inter- and intra-class shape variability while maintaining high segmentation fidelity under real-world imaging noise and illumination inconsistencies.}

\begin{table}
    \centering
    \setlength{\extrarowheight}{1.0pt}
    {
    \caption{Quantitative comparison of CATFA-Net with SOTA methods on REFUGE.}
    \resizebox{\textwidth}{!}{
    \begin{threeparttable}
    \begin{tabular}{lccccccc}
    \hline\hline
        Method & DSC (\%) & IoU (\%) & P (\%) & R (\%) & Sp (\%) & MCC (\%) & HD (mm) \\
    \hline
        U-Net \textcolor{blue}{\citep{ronneberger2015unet}} & $85.14 \pm 4.3^{*}$ & $74.61 \pm 6.5^{*}$ & $81.56 \pm 7.3^{*}$ & $91.11 \pm 4.8^{*}$ & $95.51 \pm 7.1^{*}$ & $82.60 \pm 2.1^{*}$ & $17.17 \pm 5.8^{*}$ \\
        U-Net++ \textcolor{blue}{\citep{zhou2018unet++}} & $84.47 \pm 3.5^{*}$ & $74\pm3.1^{*}$ & $80.37\pm2.4^{*}$ & $91.15\pm4.1^{*}$ & $95.55\pm5.7^{*}$ & $83.02\pm7.4^{*}$ & $15.15\pm4.6^{*}$ \\
        R2U-Net \textcolor{blue}{\citep{alom2018recurrent}} & $86.22\pm 2.2^{*}$ & $77.44\pm1.2^{*}$ & $85.33\pm6.8^{*}$ & $90.78\pm8.7^{*}$ & $96.34\pm2.5^{*}$ & $83.55\pm3^{*}$ & $16.66\pm6.1^{*}$ \\
        PSPNet \textcolor{blue}{\citep{zhao2017pyramid}} & $87.65\pm5.8^{*}$ & $77.22\pm5.7^{*}$ & $87.95\pm5.2^{\dagger}$ & $89.55\pm3.1^{*}$ & $96.15\pm6.2^{*}$ & $85.1\pm4.7^{\dagger}$ & $16.94\pm4.2^{\ddagger}$ \\
        Deeplabv3+ \textcolor{blue}{\citep{chen2018encoder}} & $86.67 \pm 7.6^{*}$ & $75.51\pm3.3^{*}$ & $90.05\pm3.9^{*}$ & $88.88\pm6.9^{*}$ & $98.13\pm4.9^{*}$ & $87.42\pm8.3^{*}$ & $16.88\pm3.7^{\dagger}$ \\
        nnUNet \textcolor{blue}{\citep{isensee2021nnu}} & $86.66 \pm 2.4^{*}$ & $75.16\pm2.8^{\dagger}$ & $82.03\pm2.6^{\dagger}$ & $92.24\pm5.9^{*}$ & $95.48\pm3.7^{\dagger}$ & $85.61\pm6.3^{\ddagger}$ & $14.66\pm7.1^{\dagger}$ \\
        AttUNet \textcolor{blue}{\citep{oktay2018attention}} & $85.11\pm8.1^{*}$ & $75.28\pm8.6^{\dagger}$ & $74.18\pm9.7^{\ddagger}$ & $93.29\pm9.2^{\dagger}$ & $93.61\pm8^{\ddagger}$ & $83.54\pm5.9^{\dagger}$ & $15.84\pm1.9^{\dagger}$ \\
        AttR2UNet \textcolor{blue}{\citep{wang2017residualatt}} & $86.58 \pm 6.4^{\dagger}$ & $76.14 \pm 6.2^{*}$ & $82.14 \pm 1.6^{*}$ & $92.55 \pm 1.4^{*}$ & $95.33 \pm 1.3^{\dagger}$ & $84.17 \pm 6.6^{*}$ & $15.45 \pm 7.6^{*}$ \\
        TransUNet \textcolor{blue}{\citep{chen2021transunet}} & $88.07 \pm 1.9^{\dagger}$ & $79.66 \pm2.9^{*}$ & $83.65 \pm 8.4^{*}$ & $\mathbf{94.34 \pm 5.6}$ & $96.22 \pm 6.5^{\ddagger}$ & $84.69 \pm 1.5^{*}$ & $13.26 \pm 3.3^{\dagger}$ \\
        SwinUNet \textcolor{blue}{\citep{cao2021swinunet}} & $89.25 \pm 5.1^{*}$ & $80.51 \pm 5.6^{\dagger}$ & $88.65 \pm 6.3^{\ddagger}$ & $91.02 \pm 2.3^{\dagger}$ & $95.12 \pm 9.6^{*}$ & $86.77 \pm 9.3^{\dagger}$ & $12.78 \pm 6.5^{*}$ \\
        UCTransNet \textcolor{blue}{\citep{UCTransNet}} & $88.78 \pm 2.7^{\dagger}$ & $81.4\pm9^{\ddagger}$ & $90.55 \pm 4.5^{\dagger}$ & $90.1 \pm 6.7^{\dagger}$ & $96.87 \pm 4^{\dagger}$ & $86.11 \pm 2.9^{\ddagger}$ & $11.18 \pm 2.4^{\ddagger}$ \\
        FCT \textcolor{blue}{\citep{tragakis2023fullyconvolutionaltransformermedical}} & $88.23 \pm 3.7^{\dagger}$ & $81.55 \pm 6.1^{\ddagger}$ & $90.64 \pm 2.7^{\ddagger}$ & $89.93 \pm 7.9^{\ddagger}$ & $97.23 \pm 3.6^{\dagger}$ & $87.41 \pm 4.6^{\dagger}$ & $11.28 \pm 3.8^{\dagger}$ \\
        \hline
        CATFA-Net-S & $90.42 \pm 3.1$ & $83.1 \pm 1.7$ & $\mathbf{91.45 \pm 3.2}$ & $90.11 \pm 3.4$ & $\mathbf{98.23 \pm 2.8}$ & $\mathbf{89.26 \pm 1.7}$ & $10.42 \pm 2$ \\
        CATFA-Net-L & $\mathbf{90.55 \pm 3}$ & $\mathbf{83.35 \pm 2.6}$ & $90.16 \pm 2.9$ & $93.55 \pm 2.6$ & $97.96 \pm 1.3$ & $88.41 \pm 3.3$ & $\mathbf{10.17 \pm 2.9}$ \\
        \hline
    \end{tabular}
    \begin{tablenotes}
    \item $^*p$-value $<0.001$, based on one-tailed Wilcoxon signed-rank test against CATFA-Net-L
    \item $^\dagger p$-value $<0.01$, based on one-tailed Wilcoxon signed-rank test against CATFA-Net-L
    \item $^\ddagger p$-value $<0.05$, based on one-tailed Wilcoxon signed-rank test against CATFA-Net-L
    \end{tablenotes}
    \end{threeparttable}}}
    \label{table-5}
\end{table}

\begin{figure}[t]
    \centering
    \includegraphics[width=1.0\linewidth]{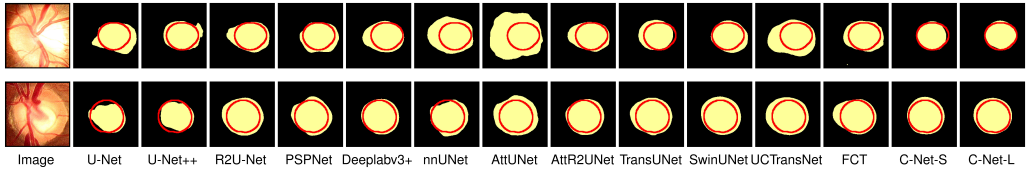}
    \caption{{\textbf{Qualitative comparison of CATFA-Net (C-Net-S and C-Net-L) variants with other SOTA methods on the REFUGE Dataset.} The red curve represents the ground-truth boundary overlaid on the segmented mask produced by each model.}}
    \label{figure-13}
\end{figure}

\subsubsection{Performance on the REFUGE dataset}
{Table 5 presents a quantitative comparison of CATFA-Net with twelve SOTA segmentation models on the REFUGE dataset. CATFA-Net-L achieves the best overall performance, obtaining the highest Dice score (90.55\%), IoU (83.35\%), and the lowest Hausdorff Distance (10.17 mm). CATFA-Net-S, also performs competitively, especially in terms of precision (91.45\%) and specificity (98.23\%), outperforming all other methods on these metrics. Compared to recent hybrid models such as UCTransNet and FCT, CATFA-Net-L delivers improved boundary accuracy and region consistency with lower standard deviation, indicating better robustness under dataset variability. Statistical testing using the one-tailed Wilcoxon signed-rank test confirms that these gains are significant, with $p < 0.01$ against most competing methods. The training and validation Dice loss curves for REFUGE (see Fig. \ref{figure-8}(e)) reveal mild divergence, which indicates overfitting. However, the high dice scores constant over $>10$ epochs (Fig. \ref{figure-8}(f)) shows the framework's effective generalization, even in the presence of challenging optic disc and cup structures with subtle contrast boundaries.}

{Qualitative comparisons in Fig. \ref{figure-13} further validate these findings. Traditional CNN-based models display noticeable shape distortions and contour misalignment, particularly in the optic cup region. Attention and transformer-based methods like AttUNet and SwinUNet offer moderate improvements in structural alignment but occasionally exhibit under-segmentation or irregular borders. In contrast, both CATFA-Net variants produce smooth, well-aligned contours that closely trace the ground truth for both the optic disc and cup.} 

\begin{table}
    \centering
    \setlength{\extrarowheight}{1.0pt}
    {
    \caption{Quantitative results of ablation studies (architectural design and attention modules) on the GLaS dataset.}
    \resizebox{\textwidth}{!}{
    \begin{threeparttable}
    \begin{tabular}{lccccccc}
    \hline\hline
    Model & DSC (\%) & IoU (\%) & P (\%) & R (\%) & Sp (\%) & MCC (\%) & HD (mm) \\
    \hline
    $\mathcal{M}_0$ & $86.21\pm6.3$ & $75.52 \pm 5.4$ & $87.53 \pm 4.7$ & $85.76 \pm 6.2$ & $88.29 \pm 5.5$ & $77.72 \pm 6.8$ & $35 \pm 4.6$ \\
    $\mathcal{N}$ & $83.54\pm7.9$ & $72.82 \pm 6.8$ & $80.45 \pm 7.2$ & $87.57 \pm 5.9$ & $77.32 \pm 6.4$ & $66.9 \pm 5.7$ & $36.85 \pm 4.4$ \\
    $\mathcal{M}_1$ & $86.95\pm5.1$ & $77.33 \pm 5.1$ & $87.94 \pm 5.9$ & $88.26 \pm 6.1$ & $89.83 \pm 4.5$ & $77.92 \pm 6.6$ & $34.68 \pm 3.2$ \\
    $\mathcal{M}_2$ & $87.76\pm4.5$ & $81.19 \pm 4.4$ & $88.12 \pm 5.3$ & $89.11 \pm 4.6$ & $90.78 \pm 3.2$ & $79.67 \pm 7.1$ & $33.46 \pm 7.4$ \\
    $\mathcal{P}_0 = \mathcal{M}_1 + \mathcal{N}$ & $91.66\pm2.8$ & $85.71 \pm 2.3$ & $91.76 \pm 2.2$ & $93.25 \pm 1.8$ & $92.23 \pm 2.1$ & $84.36 \pm 1.7$ & $30.44 \pm 5.9$ \\
    $\mathcal{P}_1$ & $93.12\pm2$ & $87.96 \pm 1.8$ & $94.27 \pm 3.3$ & $94.62 \pm 2.6$ & $93.98 \pm 2.5$ & $87.13 \pm 1.8$ & $29.64 \pm 3.1$ \\
    \hline
    CATFA-Net-L & $\mathbf{94.48 \pm 2.4}$ & $\mathbf{89.71 \pm 2.8}$ & $\mathbf{95.30 \pm 2.2}$ & $\mathbf{95.04 \pm 2.3}$ & $\mathbf{94.68 \pm 2.1}$ & $\mathbf{88.10 \pm 2.5}$ & $\mathbf{28.20 \pm 2.5}$ \\
    \hline
    \end{tabular}
    \end{threeparttable}}}
    \label{table-6}
\end{table}

\subsection{Ablation studies}
\subsubsection{Ablation results on the proposed architecture and attention modules}
{To assess the contribution of individual components in CATFA-Net, we conduct extensive ablation experiments on the GLaS dataset. Performance is evaluated using the seven segmentation metrics defined in Section 3.2, with results reported as mean $\pm$ SD over 10 independent runs in Table 6.} 

{Given that CATFA-Net is structured as a hierarchical encoder–decoder (Fig.~\ref{figure-2}), we establish two baselines for comparison. The first, denoted $\mathcal{M}_0$, is based on the Swin Transformer (Swin-T) \textcolor{blue}{\citep{liu2021swin}}, configured with a stage-wise compute ratio of $2:2:6:2$, aligning with the design of the H-CAT branch in CATFA-Net-L. The second baseline, $\mathcal{N}$, uses ConvNeXt-B \textcolor{blue}{\citep{liu2022convnet}}, which serves as the convolutional encoder backbone in CATFA-Net-L. As shown in Table 6, $\mathcal{M}_0$ achieves a Dice score of 86.21\% and Hausdorff Distance (HD) of 35.0 mm, while $\mathcal{N}$ reaches 83.54\% Dice and 36.85 mm HD—demonstrating the relative strength of transformer-based encoders in capturing long-range dependencies in histological textures.}

{\textbf{Effectiveness of the CAT Block.}
We first replace the W-MSA and MLP components in the Swin-T block with the proposed context addition self-attention (CASA) and depthwise fully convolutional network (d-FCN) modules, forming the Context Addition Transformer (CAT) block (Fig.~\ref{figure-3}(a); Section 2.1). Incorporating these blocks into $\mathcal{M}_0$ yields variant $\mathcal{M}_1$, which improves the mean Dice score by 0.74\%, and HD reduction by 0.32 mm, indicating that CAT blocks better capture inter-image similarity and positional cues. Note that $\mathcal{M}_1$ closely resembles the H-CAT branch of CATFA-Net-L but still does not include the proposed CCTFA module, which plays a crucial role in trans-branch information exchange of the proposed architecture.}

{\textbf{Impact of the H-CAT Encoder branch.} The CCTFA module consists of two sub-components: (1) a spatial attention mechanism applied exclusively to ConvNeXt outputs, and (2) a cross-channel attention mechanism that jointly processes outputs from both the CAT and ConvNeXt branches using separate pathways (see Fig. \ref{figure-5} and Section 2.3). To isolate the impact of the spatial attention pathway and to construct a standalone H-CAT encoder, we create model $\mathcal{M}_2$ by disabling the spatial attention sub-module and retaining only the CAT block driven pathways of the cross-channel attention sub module, which leads to further 0.81\% gain in Dice score, and reduction of 1.22 mm in HD over $\mathcal{M}_1$. This improvement validates the role of CAT blocks in preserving global context during hierarchical downsampling and demonstrates their standalone efficacy in segmentation tasks.} 

{\textbf{Effectiveness of a trans-convolutional architecture.} To validate the impact of trans-convolutional design, we combine $\mathcal{M}_1$ with the ConvNeXt-B encoder ($\mathcal{N}$) using a simple fusion strategy (feature concatenation followed by $1 \times 1$ convolution). This configuration, denoted as $\mathcal{P}_0$, achieves a 5.45\% Dice improvement over $\mathcal{M}_0$, 8.12\% over $\mathcal{N}$, 4.71\% over $\mathcal{M}_1$, and 3.9\% over $\mathcal{M}_2$. These gains demonstrate the synergistic benefit of fusing convolutional and transformer-based features and underscore the overall strength of a trans-convolutional hybrid framework for robust medical image segmentation.}

{\textbf{Effectiveness of the CCTFA and SAFG attention modules.} We next assess the benefit of the `proposed' trans-convolutional design by replacing the standard fusion module with the proposed CCTFA attention sub module, creating the variant $\mathcal{P}_1$. This results in a DSC of 93.12\%, and HD of 29.64 mm, which highlights the effectiveness of joint spatial and channel-wise attention in capturing complementary features from both encoder branches. Finally, incorporating the Spatial Attention Fusion Gate (SAFG) module into $\mathcal{P}_1$ completes the CATFA-Net-L architecture, which achieves a Dice score of 93.30\%, establishing a new SOTA for this benchmark and confirming the cumulative benefits of each architectural enhancement.}

\begin{table}
    \centering
    \setlength{\extrarowheight}{1.0pt}
    {
    \caption{Quantative results of ablation studies (encoder configuraton) on the GLaS Dataset}
    \resizebox{\textwidth}{!}{
    \begin{threeparttable}    
    \begin{tabular}{lcccccccc}
    \hline\hline
    Model & Encoder configuration & DSC (\%) & IoU (\%) & P (\%) & R (\%) & Sp (\%) & MCC (\%) & HD (mm) \\
    \hline
    $\mathcal{S}_0$ & $\text{Res}18 + \text{Swin-T}$ & $86.13 \pm 7.2$ & $79.92 \pm 6.4$ & $91.07 \pm 9.1$ & $91.02 \pm 5.8$ & $90.45 \pm 6.6$ & $82.09 \pm 8.9$ & $35.77 \pm 6.0$\\
    $\mathcal{S}_1$ & $\text{Res}18 + \text{H-CAT}$ & $87.47 \pm 5.6$ & $83.01 \pm 7.1$ & $91.19 \pm 6.2$ & $90.84 \pm 4.1$ & $92.06 \pm 7.4$ & $85.63 \pm 6.3$ & $33.68 \pm 6.6$\\
    $\mathcal{S}_2$ & $\text{ConvNeXt-T} + \text{Swin-T}$ & $90.89 \pm 7.0$ & $85.51 \pm 8.4$ & $91.42 \pm 5.7$ & $92.63 \pm 4.3$ & $92.58 \pm 6.0$ & $85.87 \pm 5.1$ & $31.69 \pm 9.2$\\
    CATFA-Net-S & $\text{ConvNeXt-T} + \text{H-CAT}$ & $\mathbf{94.04 \pm 1.9}$ & $\mathbf{87.9 \pm 1.8}$ & $\mathbf{94.16 \pm 3.3}$ & $\mathbf{94.56 \pm 2.6}$ & $\mathbf{94.6 \pm 2.5}$ & $\mathbf{87.1 \pm 1.8}$ & $\mathbf{29.64 \pm 3.1}$\\
    \hline
    $\mathcal{L}_0$ & $\text{Res}34 + \text{Swin-T}$ & $88.53 \pm 6.8$ & $84.51 \pm 5.3$ & $90.07 \pm 7.2$ & $90.78 \pm 6.0$ & $91.99 \pm 4.6$ & $84.04 \pm 7.7$ & $33.56 \pm 8.4$\\
    $\mathcal{L}_1$ & $\text{Res}34 + \text{H-CAT}$ & $90.12 \pm 4.9$ & $85.18 \pm 6.6$ & $91.06 \pm 5.5$ & $92.64 \pm 8.2$ & $93.03 \pm 6.3$ & $83.08 \pm 7.5$ & $31.32 \pm 6.4$\\
    $\mathcal{L}_2$ & $\text{ConvNeXt-B} + \text{Swin-T}$ & $92.06 \pm 8.1$ & $87.35 \pm 4.3$ & $93.67 \pm 7.6$ & $93.08 \pm 6.1$ & $93.09 \pm 9.0$ & $86.02 \pm 6.5$ & $29.71 \pm 5.7$\\
    CATFA-Net-L & $\text{ConvNeXt-B} + \text{H-CAT}$ & $\mathbf{94.48 \pm 2.4}$ & $\mathbf{89.71 \pm 2.8}$ & $\mathbf{95.30 \pm 2.2}$ & $\mathbf{95.04 \pm 2.3}$ & $\mathbf{94.68 \pm 2.1}$ & $\mathbf{88.10 \pm 2.5}$ & $\mathbf{28.20 \pm 2.5}$ \\
    \hline
    \end{tabular}
    \end{threeparttable}}}
    \label{table-7}
\end{table}

\subsubsection{Ablation results on the encoder configurations}
{In this second set of ablation experiments, we investigate the impact of different encoder configurations within CATFA-Net variants on the GLaS dataset. All models employ the lightweight Conv-G-NeXt decoder and include the proposed CCTFA and SAFG modules to ensure consistent feature fusion across comparisons.}

{We begin with the CATFA-Net-S architecture and evaluate a baseline configuration, denoted as $\mathcal{S}_0$, which uses a dual-encoder setup comprising ResNet-18 (pretrained on ImageNet-1K) and Swin-T as the convolutional and transformer encoders, respectively. Replacing the Swin-T stages with the proposed CAT blocks while retaining ResNet-18 as the convolutional encoder yields model $\mathcal{S}_1$, which improves the Dice score by 1.34\% over $\mathcal{S}_0$. This gain highlights the benefit of introducing global contextual interactions through the CAT block’s neighboring key-query attention mechanism, particularly advantageous in capturing the morphological variability common in medical images. Next, we evaluate the impact of ConvNeXt-T \textcolor{blue}{\citep{liu2022convnet}} as a convolutional encoder. Substituting ResNet-18 with ConvNeXt-T (while retaining Swin-T) produces model $\mathcal{S}_2$, resulting in a further 3.42\% improvement in Dice score over $\mathcal{S}_1$. This is attributed to the superior representational efficiency of ConvNeXt, which preserves long-range spatial information through hierarchical depthwise convolutions. Replacing Swin-T with CAT blocks in this configuration restores the full CATFA-Net-S encoder design, achieving an additional 3.15\% gain and confirming the advantage of the proposed trans-convolutional configuration.}

{Parallel experiments were conducted for CATFA-Net-L, using ResNet-34 and ConvNeXt-B as the convolutional backbones, with stage compute ratios scaled accordingly. As summarized in Table 7, these results demonstrate consistent improvements across variants and validate the architectural choices underlying both CATFA-Net-S and CATFA-Net-L.}

\begin{figure}
    \centering
    \includegraphics[width=1\linewidth]{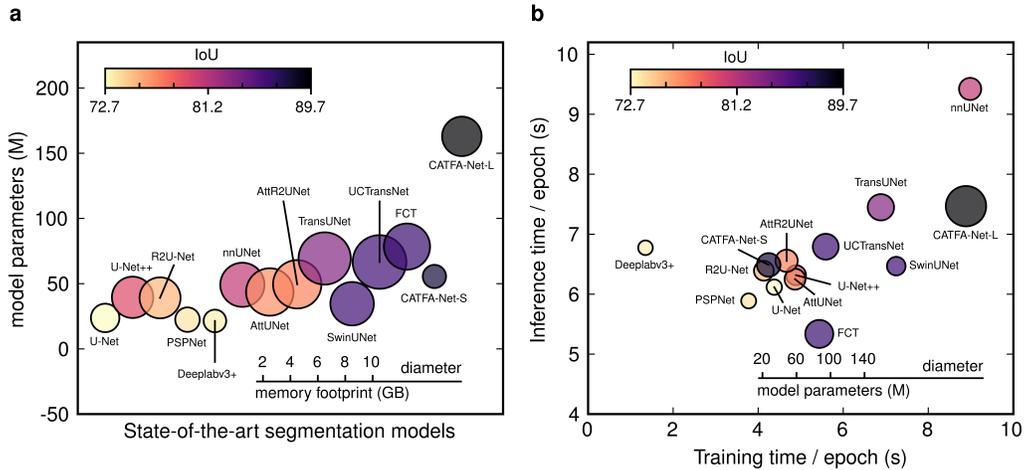}
    \caption{{\textbf{Computational complexity comparison of SOTA segmentation models evaluated on the GLaS Dataset. (a)} Comparison of model parameters (in millions) vs. memory footprint (in GB) for various segmentation models. Each circle represents a model, with its diameter proportional to memory consumption, vertical position indicating parameter count, and color representing mean IoU performance on the GLaS dataset. \textbf{(b)} Comparison of training time per epoch (x-axis) versus inference time per epoch (y-axis), with circle size indicating model parameters and color again denoting mean IoU.}}
    \label{figure-14}
\end{figure}

\subsection{Computational complexity and runtime efficiency}
{To assess the feasibility of our proposed framework in real-world clinical applications, we provide a comprehensive analysis of computational efficiency, including parameter count, memory footprint, training time, and inference time. This evaluation is performed for both CATFA-Net-S, CATFA-Net-L and a diverse set of recent SOTA segmentation models (see section 3.5) on the GLaS dataset. These metrics are particularly important for practical deployment in time-sensitive clinical workflows and on embedded platforms with limited computational resources. To ensure fairness, all experiments were conducted on a single NVIDIA Tesla V100 GPU with 32 GB of memory, using identical training and evaluation configurations (see section 3.2).}

{As depicted in Fig.~\ref{figure-14}(a), classical convolutional models such as U-Net, PSPNet, and DeepLabv3+ exhibit low GPU memory usage and compact model sizes ($<$ 30M parameters), but their limited capacity to model long-range dependencies results in suboptimal IoU scores. More advanced convolutional networks like nnUNet, U-Net++, and R2U-Net improve segmentation accuracy but introduce significantly larger parameter counts ($>30$M) and increased memory consumption. Transformer-based architectures such as TransUNet offer improved performance via global self-attention but require heavy computational overhead, with memory footprints frequently exceeding 6–7 GB and parameter counts surpassing 50M. Similarly, trans-convolutional hybrids like FCT and UCTransNet, while achieving high accuracy, demand substantial resources for both training and inference. In contrast, CATFA-Net-S achieves superior IoU scores with a markedly lower memory footprint, attributed to efficient downsampling in early encoder stages and a lightweight decoder, which comprises only 10\% of the total model parameters. Notably, CATFA-Net-S outperforms UCTransNet and FCT by $\sim$2\% and 1.6\% in IoU, respectively, while consuming $3.7\times$ and $2.6\times$ less GPU memory—highlighting the strength of its trans-convolutional design for real-world use cases. Meanwhile, CATFA-Net-L sets a new benchmark, achieving an IoU of 89.71\% with $1.5\times$ lower GPU memory usage compared to UCTransNet, albeit having a parameter count of $>150$M.}

{Fig.~\ref{figure-14}(b) extends this analysis by comparing training and inference time per epoch across models, averaged over 50 iterations from 10 independent runs. Architectures following the U-shaped encoder–decoder paradigm with heavy convolutional backbones—such as U-Net, AttUNet, R2U-Net, and U-Net++, show relatively balanced training (4–6s/epoch) and inference times (6–7s/epoch). Interestingly, DeepLabv3+, despite having a comparable inference time, exhibits faster training (1–2s/epoch) due to its lightweight decoder design, which results in lower backpropagation cost and fewer parameter updates in the segmentation head. Conversely, nnUNet incurs higher training and testing time per epoch, largely due to its extensive data preprocessing and augmentation pipeline. Transformer-based models (e.g., TransUNet, SwinUNet) demonstrate high segmentation performance but suffer from longer runtimes due to the quadratic complexity of self-attention and large model sizes with large token embeddings. Hybrid models like UCTransNet strike a balance, reducing training time by up to 1.5× compared to full-transformer models, while preserving accuracy. CATFA-Net-S benefits from a similar hybrid strategy: its hierarchical structure, efficient attention design, and lightweight decoder together reduce training time to $\sim4$s/epoch without compromising inference speed or segmentation quality. CATFA-Net-L, while achieving the best performance, incurs a higher training time of $\sim9$s/epoch and inference time of $\sim7.5$s/epoch due to its deeper encoder configuration and increased parameter count.}

{These findings demonstrate that CATFA-Net-S offers a favorable trade-off between accuracy and efficiency, making it a practical and scalable solution for high-precision medical image segmentation in resource-constrained clinical settings.}

\begin{table}[t]
    \centering
    \setlength{\extrarowheight}{1.0pt}
    {
    \caption{Quantitative comparison of cross-dataset generalization performance.}
    \resizebox{\textwidth}{!}{
    \begin{threeparttable}
    \begin{tabular}{lccccccc}
    \hline\hline
        Method & DSC (\%) & IoU (\%) & P (\%) & R (\%) & Sp (\%) & MCC (\%) & HD (mm) \\
        \hline
        U-Net \textcolor{blue}{\citep{ronneberger2015unet}}& $71.11 \pm 7.6^{\ddagger}$ & $53.67 \pm 4.4^{\ddagger}$ & $66.39 \pm 5.1^{\ddagger}$ & $73.54 \pm 3.9^{\ddagger}$ & $94.82 \pm 6.5$ & $65.26 \pm 2.8^{\ddagger}$ & $32.88 \pm 8.9^{\ddagger}$ \\
        U-Net++ \textcolor{blue}{\citep{zhou2018unet++}}& $73.19 \pm 2.3^{\ddagger}$ & $56.14 \pm 5.9^{\ddagger}$ & $67.33 \pm 6.3$ & $75.5 \pm 2.9$ & $94.13 \pm 7.7^{\ddagger}$ & $67.78 \pm 4.4^{\ddagger}$ & $30.35 \pm 7.1^{\ddagger}$ \\
        R2U-Net \textcolor{blue}{\citep{alom2018recurrent}} & $67.61 \pm 8.1^{\ddagger}$ & $52.31 \pm 4.6^{\ddagger}$ & $62.53 \pm 3.5^{\ddagger}$ & $77.15 \pm 7.3$ & $93.07 \pm 9.2^{\ddagger}$ & $62.18 \pm 3.3^{\ddagger}$ & $34.56 \pm 4.9^{\ddagger}$ \\
        PSPNet \textcolor{blue}{\citep{zhao2017pyramid}} & $60.35 \pm 3.4^{\ddagger}$ & $44.29 \pm 9.1^{\ddagger}$ & $55.53 \pm 5.6^{\ddagger}$ & $65.66 \pm 3.1^{\ddagger}$ & $93.62 \pm 7.4^{\ddagger}$ & $54.48 \pm 2.7^{\ddagger}$ & $34.07 \pm 4.8^{\ddagger}$ \\
        DeepLabV3+ \textcolor{blue}{\citep{chen2018encoder}} & $65.19 \pm 4.5^{\ddagger}$ & $50.31 \pm 6.9^{\ddagger}$ & $67.06 \pm 5.7^{\ddagger}$ & $63.26 \pm 3.2^{\ddagger}$ & $95.32 \pm 8^{\ddagger}$ & $61.2 \pm 6.7^{\ddagger}$ & $38.84 \pm 4.1^{\ddagger}$ \\
        nnUNet \textcolor{blue}{\citep{isensee2021nnu}} & $69.18 \pm 3.5^{\ddagger}$ & $54.19 \pm 6.4^{\ddagger}$ & $65.55 \pm 3.2^{\ddagger}$ & $73.62 \pm 5.9^{\ddagger}$ & $94.76 \pm 5.5^{\ddagger}$ & $65.31 \pm 3.9^{\ddagger}$ & $31.56 \pm 6.6^{\ddagger}$ \\
        AttUNet \textcolor{blue}{\citep{oktay2018attention}} & $68.52 \pm 5.2^{\ddagger}$ & $54.41 \pm 2.9^{\ddagger}$ & $66.46 \pm 6.8^{\ddagger}$ & $72.62 \pm 9.3^{\ddagger}$ & $93.55 \pm 2.7^{\ddagger}$ & $64.49 \pm 8.2^{\ddagger}$ & $35.12 \pm 5.3^{\ddagger}$ \\
        AttR2UNet \textcolor{blue}{\citep{wang2017residualatt}} & $67.48 \pm 6.4^{\ddagger}$ & $51.62 \pm 9.1^{\ddagger}$ & $64.36 \pm 7.9^{\ddagger}$ & $74.29 \pm 4.5^{\ddagger}$ & $95 \pm 6.6$ & $63.23 \pm 3.7^{\ddagger}$ & $35.94 \pm 4.0^{\ddagger}$ \\
        TransUNet \textcolor{blue}{\citep{chen2021transunet}}& $71.47 \pm 4.2^{\ddagger}$ & $57.43 \pm 8.5^{\ddagger}$ & $68.32 \pm 2.1$ & $77.69 \pm 5.4$ & $94.57 \pm 3.6^{\ddagger}$ & $69.46 \pm 6.1^{\ddagger}$ & $27.21 \pm 9.0^{\ddagger}$ \\
        SwinUNet \textcolor{blue}{\citep{cao2021swinunet}} & $74.57 \pm 5.8^{\ddagger}$ & $61.32 \pm 9.4^{\ddagger}$ & $68.67 \pm 3.9$ & $\mathbf{83.12 \pm 8.1}$ & $95.35 \pm 6.3$ & $72.81 \pm 2.6^{\ddagger}$ & $28.6 \pm 3.7^{\ddagger}$ \\
        UCTransNet \textcolor{blue}{\citep{UCTransNet}} & $76.39 \pm 6.5$ & $63.08 \pm 4.9^{\ddagger}$ & $74.16 \pm 3.6$ & $77.55 \pm 7.8$ & $95.63 \pm 8.4$ & $72.48 \pm 6.9^{\ddagger}$ & $27.81 \pm 5.5^{\ddagger}$ \\
        FCT \textcolor{blue}{\citep{tragakis2023fullyconvolutionaltransformermedical}} & $76.67 \pm 4.3$ & $61.08 \pm 3.3^{\ddagger}$ & $74.68 \pm 2.6$ & $76.43 \pm 6.4$ & $94.56 \pm 5.1^{\ddagger}$ & $73.16 \pm 4.6^{\ddagger}$ & $25.61 \pm 5.3$ \\ 
        \hline
        CATFA-Net-S & $77.47 \pm 1.9$ & $64.56 \pm 2.5$ & $75.89 \pm 1.4$ & $78.53 \pm 2.2$ & $96.02 \pm 2.9$ & $74.22 \pm 1.8$ & $\mathbf{24.69 \pm 1.6}$ \\
        CATFA-Net-L & $\mathbf{78.49 \pm 1.5}$ & $\mathbf{64.75 \pm 1.2}$ & $\mathbf{78.88 \pm 2.0}$ & $79.16 \pm 2.8$ & $\mathbf{96.34 \pm 1.9}$ & $\mathbf{75.24 \pm 1.7}$ & $\mathbf{24.92 \pm 2.3}$ \\
    \hline
    \end{tabular}
    \begin{tablenotes}
    \item $^{\ddagger} p$-value $<0.05$, based on one-tailed Wilcoxon signed-rank test against CATFA-Net-L
    \end{tablenotes}
    \end{threeparttable}}}
\end{table}

\subsection{Robustness evaluation}
\subsubsection{Cross-dataset generalization}
{To assess the generalization ability of the proposed framework, we conduct an external validation experiment by training all models on the GLaS dataset and evaluating them on the unseen TNBC dataset. This setting tests the model’s ability to transfer learned representations across significant domain shifts, reflecting real-world challenges in deploying segmentation models across heterogeneous medical imaging sources. As summarized in Table 8, CATFA-Net-L achieves the highest performance across nearly all metrics, including a Dice score of $78.49\%$, IoU of $64.75\%$, MCC of $75.42\%$, and a Hausdorff Distance of 24.92 mm. While SwinUNet attains slightly better recall, CATFA-Net-L consistently delivers a more balanced performance with superior precision and lower boundary error. These results highlight CATFA-Net’s ability to model both fine-grained nuclear morphology and broader spatial structures in the presence of distributional shift, outperforming transformer-heavy and convolutional baselines alike.} 

\begin{table}
    \centering
    \setlength{\extrarowheight}{1.0pt}
    {
    \caption{Quantitative comparison of SOTA attention model on GLaS}
    \resizebox{\textwidth}{!}{
    \begin{threeparttable}
    \begin{tabular}{lccccccc}
    \hline\hline
    Method & DSC (\%) & IoU (\%) & P (\%) & R (\%) & Sp (\%) & MCC (\%) & HD (mm) \\
    \hline
        AG-Net \textcolor{blue}{\citep{shi2022agnet}} & $90.22 \pm 5.9^{\dagger}$ & $83.09\pm7.1^{\dagger}$ & $88.84 \pm 6.4^{\dagger}$ & $91.45 \pm 4.3^{\ddagger}$ & $89.07 \pm 8.7^{\ddagger}$ & $81.68 \pm 6.2^{\dagger}$ & $33.21 \pm 7.8$\\
        SANet \textcolor{blue}{\citep{zhong2020squeeze}} & $90.42 \pm 5.5^{\dagger}$ & $82.56 \pm 6.1^{\ddagger}$ & $88.96 \pm 8.0^{\dagger}$ & $91.59 \pm 4.7^{\dagger}$ & $89.75 \pm 5.6^{\dagger}$ & $86.02 \pm 7.5$ & $32.14 \pm 7.3$ \\
        SENet \textcolor{blue}{\citep{hu2018squeeze}} & $87.45 \pm 6.6^{\dagger}$ & $79.85 \pm 5.3^{\ddagger}$ & $84.5 \pm 7.1^{\dagger}$ & $90.84 \pm 8.6^{\dagger}$ & $85.31\pm5.4^{\dagger}$ & $75.96 \pm 7.2^{\dagger}$ & $35.33 \pm 6.8^{\dagger}$\\
        scSENet \textcolor{blue}{\citep{roy2018concurrent}} & $88.89 \pm 4.9^{\dagger}$ & $80.5 \pm 6.8^{\dagger}$ & $85.7 \pm 6.1^{\dagger}$ & $91.5 \pm 5.2^{\dagger}$ & $84.7 \pm 9.3^{\ddagger}$ & $79.11 \pm 7.7^{\ddagger}$ & $34.14 \pm 8.8^{\ddagger}$\\
        AAU-Net \textcolor{blue}{\citep{chen2022aau}} & $91.88 \pm 6.3$ & $85.35 \pm 6.6^{\ddagger}$ & $91.57 \pm 4.6^{\dagger}$ & $90.89 \pm 5.5^{\dagger}$ & $90.53 \pm 8.2^{\ddagger}$ & $83.98 \pm 4.8^{\ddagger}$ & $30.1 \pm 9.1$\\
        \hline
        CATFA-Net-S & $94.04 \pm 1.9$ & $87.90 \pm 1.8$ & $94.16 \pm 3.3$ & $94.56 \pm 2.6$ & $94.6 \pm 2.5$ & $87.10 \pm 1.8$ & $29.64 \pm 3.1$ \\
        CATFA-Net-L & $\mathbf{94.48 \pm 2.4}$ & $\mathbf{89.71 \pm 2.8}$ & $\mathbf{95.30 \pm 2.2}$ & $\mathbf{95.04 \pm 2.3}$ & $\mathbf{94.68 \pm 2.1}$ & $\mathbf{88.10 \pm 2.5}$ & $\mathbf{28.20 \pm 2.5}$ \\
        \hline
    \hline
     \end{tabular}
    \begin{tablenotes}
    \item $^\dagger p$-value $<0.01$, based on one-tailed Wilcoxon signed-rank test against CATFA-Net-L
    \item $^\ddagger p$-value $<0.05$, based on one-tailed Wilcoxon signed-rank test against CATFA-Net-L
    \end{tablenotes}
     \end{threeparttable}}}
     \label{table-9}
\end{table}

\subsubsection{Comparison with SOTA attention mechanisms}
{To further evaluate the effectiveness of CATFA-Net's attention design, we compare it against several SOTA attention-based segmentation models, including AG-Net \textcolor{blue}{\citep{shi2022agnet}}, SANet \textcolor{blue}{\citep{zhong2020squeeze}}, SENet \textcolor{blue}{\citep{hu2018squeeze}}, scSE-Net \textcolor{blue}{\citep{roy2018concurrent}}, and AAU-Net \textcolor{blue}{\citep{chen2022aau}}—each incorporating spatial, channel-wise, or self-attention mechanisms. All models are trained and evaluated on the GLaS dataset using the protocol described in Section 3.2. As shown in Table 9, CATFA-Net-L achieves the highest performance across all evaluation metrics, including a Dice score of 94.48\%, IoU of 89.71\%, MCC of 88.1\%, and the lowest Hausdorff Distance of 28.2 mm. This performance gain can be attributed to CATFA-Net’s staged dual-attention strategy, which unifies cross-channel transformer attention with spatial refinement following feature fusion. Unlike traditional attention models that often focus on a single modality of attention, CATFA-Net is better equipped to model the intricate anatomical structures and heterogeneous textures found in medical images. Additionally, the integration of spatial attention fusion in the decoder stages enables effective suppression of upsampling-induced noise, resulting in improved robustness and generalization across diverse segmentation challenges.}

\section{Discussions}
{This work introduces CATFA-Net, a hybrid hierarchical trans-convolutional architecture that synergistically combines the strengths of convolutional and transformer-based representations for medical image segmentation. At its core, CATFA-Net employs a dual-branch encoder: a ConvNeXt-based convolutional branch adept at capturing fine-grained local spatial features, and a Hierarchical Cross-Attention Transformer (H-CAT) branch tailored to model long-range contextual dependencies. The H-CAT encoder incorporates two key innovations—Context Addition Self-Attention and a depth-wise fully convolutional module—enabling effective inter-image resemblance modeling while preserving positional fidelity. These branches are fused via the Cross Channel Trans-Convolutional Fusion Attention (CCTFA) module, which adaptively merges global and local cues into rich, discriminative representations. The decoder further refines these features using a Spatial Attention Fusion Gate (SAFG), which selectively enhances salient structures while suppressing background noise. Together, these components form a powerful end-to-end architecture, enabling CATFA-Net to consistently outperform existing state-of-the-art segmentation models across multiple medical imaging benchmarks. In the remainder of this section, we analyze the contributions of individual components, interpret empirical results, and discuss broader implications of the proposed design, including its generalizability across datasets.}

{CATFA-Net exhibits consistent superiority across five diverse datasets (GLaS, DSB 2018, CVC-ClinicDB, ISIC 2018, REFUGE), showing notable improvements in critical segmentation metrics such as Dice Similarity Coefficient (DSC) and Hausdorff Distance (HD) (Sections 3.3.1–3.3.5, Tables 1–5). These improvements validate our hypothesis that combining local and global modeling capacities results in superior performance, particularly in complex or noisy anatomical settings. CATFA-Net establishes new SOTA results on GLaS and ISIC 2018, while matching top-tier performance on CVC-ClinicDB. These gains stem from its dual-encoder structure and cohesive attention mechanisms (CCTFA and SAFG), which are absent in most U-Net-like baselines. Importantly, CATFA-Net also generalizes well across domains. In external validation experiments (Section 3.6, Table 8), it demonstrates significantly improved robustness on the TNBC dataset after being trained on DSB 2018. This underscores its potential for deployment in real-world clinical settings with unseen data distributions.}

{Convolutional networks, with their local receptive fields and strong spatial priors, are effective at capturing fine structural details but often fail to model long-range dependencies, leading to spatially fragmented outputs in complex cases (Fig. \ref{figure-1}) \textcolor{blue}{\citep{ronneberger2015unet,zhou2018unet++,isensee2021nnu}}. Transformer-based models overcome this by modeling global relationships through self-attention but often at the cost of higher computational complexity and weaker positional encoding \textcolor{blue}{\citep{chen2021transunet,liu2021swin,cao2021swinunet,xie2021segformer}}. CATFA-Net bridges this gap through its dual-branch encoder. Ablation studies (Section 3.4, Tables 6–7) confirm that the hybrid design yields substantial performance gains over single-branch models. The CCTFA module plays a central role in enabling effective fusion of convolutional and transformer features, while the SAFG module in the decoder further sharpens segmentation boundaries by enhancing spatially relevant foreground structures (Fig. \ref{figure-7}). Comparisons with recent attention-enhanced architectures (e.g., AG-Net, SANet, AAU-Net) in Table 9 demonstrate that CATFA-Net achieves both higher segmentation accuracy and better cross-domain generalization, indicating strong robustness to domain shifts.}

{Beyond accuracy, computational efficiency is critical for clinical adoption. CATFA-Net-S delivers state-of-the-art performance with a markedly smaller memory footprint and reduced inference time compared to transformer-heavy models (e.g., UCTransNet, FCT), as shown in Section 3.5 (Fig. \ref{figure-14}). This efficiency is driven by its hierarchical design, efficient attention formulations, and lightweight decoder. Meanwhile, the larger CATFA-Net-L variant, though more computationally demanding, achieves peak segmentation performance and serves as a strong upper bound for deployment scenarios with ample hardware.}

{Despite these advantages, CATFA-Net has limitations. The larger variant incurs high computational costs, making it less suitable for low-resource environments. Like most deep learning models, CATFA-Net depends on annotated datasets, which are often scarce in medical imaging. Although our cross-dataset generalization experiments reveal improved robustness compared to existing methods, performance degradation still occurs under substantial domain shifts—highlighting a persistent challenge in clinical translation.}

{Future work could focus on reducing annotation dependency through self-supervised pretraining, contrastive learning, or semi-supervised methods. Extending CATFA-Net to multi-task learning (e.g., segmentation + classification) and domain adaptation frameworks may enhance its flexibility across clinical contexts. Incorporating attention visualizations may help interpret model decisions, thereby improving clinician trust. Finally, model compression and more efficient attention variants could facilitate real-time deployment on resource-constrained platforms.}

\section{Conclusion}
This study presents CATFA-Net, a novel and efficient segmentation framework that addresses the limitations of existing hybrid models by incorporating a hierarchical trans-convolutional encoder architecture and a lightweight convolutional decoder backbone. By leveraging the Context Addition Attention mechanism and Cross-Channel Attention fusion, CATFA-Net effectively captures long-range dependencies while maintaining computational efficiency and robust performance across various datasets. The results from comprehensive evaluations on five publicly available datasets demonstrate CATFA-Net’s superiority in segmentation accuracy, achieving new state-of-the-art Dice scores applied on GLaS and ISIC 2018. Its strong generalization capabilities demonstrate through robustness analyses and external validation, further highlight its potential as a leading architecture for medical image segmentation tasks. CATFA-Net strikes a balance between performance and efficiency, providing a promising solution for high-quality segmentation in real-world applications.

\section*{CrediT authorship contribution statement}
\textbf{Siddhartha Mallick:} Writing–review \& editing, Writing–original draft, Visualization, Validation, Software, Resources, Methodology, Investigation, Formal analysis, Data curation, Conceptualization. \textbf{Aayushman Ghosh:} Writing–review \& editing, Writing–original draft, Visualization, Validation, Software, Resources, Methodology, Investigation, Formal analysis, Data curation, Conceptualization. \textbf{Jayanta Paul:} Writing–review \& editing, Supervision, Methodology, Conceptualization. \textbf{Jaya Sil:} Writing–review \& editing, Supervision, Project administration.

\section*{Declaration of competing interest}
The authors declare that they have no known competing financial interests or personal relationships that could have appeared to influence the work reported in this paper.

\section*{Acknowledgments}
This research did not receive any specific grant from funding agencies in the public, commercial, or not-for-profit sectors.

\section*{Data availability Statements}
The datasets utilized in this study are publicly available. The source code has also been made publicly accessible.

%% References
\bibliographystyle{elsarticle-harv} 
\bibliography{ref}

\end{document}